\documentclass[twocolumn,appendixfloats,tighten]{aastex631}
\usepackage{graphicx}	
\usepackage{amsmath}	
\usepackage{amssymb}	
\usepackage{units}
\usepackage{tabu}
\usepackage{multirow}
\usepackage{enumitem}
\usepackage{bm}
\usepackage{color}
\usepackage{boldline}
\usepackage{booktabs}
\usepackage{float}

\newcommand{\kms}{\, {\rm km\, s}^{-1}}

\newcommand{\be}{\begin{equation}}
\newcommand{\ee}{\end{equation}}

\def\h2{${\rm\,H_2}$}

\newcommand{\co}{\ensuremath{\mathrm{CO}\,(3-2)}}

\def\code#1{\texttt{#1}}
\def\edit#1{\textcolor{orange}{#1}} 

\newcommand{\zA}{0.3212}
\newcommand{\RAA}{$21h44m25.255s$}
\newcommand{\DecA}{$-40^{\circ}54'00.10''$}
\newcommand{\RAerrA}{$0.015s$}
\newcommand{\DecerrA}{$0.176''$}
\newcommand{\hostRAA}{$21h44m25.256s$}
\newcommand{\hostDecA}{$-40^{\circ}54'00.8''$}
\newcommand{\obsdateA}{2019-11-26}
\newcommand{\metA}{$0.03^{+0.08}_{-0.08}$} 
\newcommand{\SFRA}{$0.62^{+0.32}_{-0.24}$} 
\newcommand{\massA}{$10.39^{+0.02}_{-0.02}$} 
\newcommand{\beammajA}{$1.43''$}
\newcommand{\beamminA}{$1.10''$}
\newcommand{\beamPAA}{$83.9^{\circ}$}

\newcommand{\zB}{0.4755} 
\newcommand{\RAB}{$21h49m23.63s$}
\newcommand{\DecB}{$-52^{\circ}58'15.4''$}
\newcommand{\RAerrB}{$0.43s$}
\newcommand{\DecerrB}{$2.4''$}
\newcommand{\hostRAB}{$21h49m23.660s$}
\newcommand{\hostDecB}{$-52^{\circ}58'15.28''$}
\newcommand{\obsdateB}{2019-11-26}
\newcommand{\metB}{$-0.17^{+0.12}_{-0.12}$}
\newcommand{\SFRB}{$1.54^{+0.99}_{-0.65}$}
\newcommand{\massB}{$9.87^{+0.07}_{-0.07}$}
\newcommand{\beammajB}{$1.54''$}
\newcommand{\beamminB}{$1.39''$}
\newcommand{\beamPAB}{$87.2^{\circ}$}

\newcommand{\zC}{0.2912} 
\newcommand{\RAC}{$21h29m39.76s$}
\newcommand{\DecC}{$-79^{\circ}28'32.5''$}
\newcommand{\RAerrC}{$0.30s$}
\newcommand{\DecerrC}{$1.0''$}
\newcommand{\hostRAC}{$21h29m39.577s$}
\newcommand{\hostDecC}{$-79^{\circ}28'32.52''$}
\newcommand{\obsdateC}{2019-11-30}
\newcommand{\metC}{$-0.51^{+0.78}_{-0.51}$}
\newcommand{\SFRC}{$0.40^{+0.31}_{-0.11}$}
\newcommand{\massC}{$9.69^{+0.09}_{-0.11}$}
\newcommand{\beammajC}{$1.95''$}
\newcommand{\beamminC}{$1.37''$}
\newcommand{\beamPAC}{$-6.2^{\circ}$}

\newcommand{\zD}{0.2365} 
\newcommand{\RAD}{$12h15m55.13s$}
\newcommand{\DecD}{$-13^{\circ}01'15.6''$}
\newcommand{\RAerrD}{$0.038s$}
\newcommand{\DecerrD}{$0.4''$}
\newcommand{\hostRAD}{$12h15m55.090s$}
\newcommand{\hostDecD}{$-13^{\circ}01'15.96''$}
\newcommand{\obsdateD}{2022-04-08}
\newcommand{\metD}{$0.11^{+0.22}_{-0.23}$}
\newcommand{\SFRD}{$1.93^{+1.08}_{-0.69}$}
\newcommand{\massD}{$10.22^{+0.04}_{-0.04}$}
\newcommand{\beammajD}{$1.38''$}
\newcommand{\beamminD}{$0.91''$}
\newcommand{\beamPAD}{$-77.2^{\circ}$}

\newcommand{\zE}{0.161} 
\newcommand{\RAE}{$15h18m49.54s$}
\newcommand{\DecE}{$+12^{\circ}22'36.3''$}
\newcommand{\RAerrE}{$0.021s$}
\newcommand{\DecerrE}{$1.01''$}
\newcommand{\hostRAE}{$15h18m49.520s$}
\newcommand{\hostDecE}{$+12^{\circ}22'35.80''$}
\newcommand{\obsdateE}{2022-03-11}
\newcommand{\metE}{$-0.12^{+0.06}_{-0.06}$}
\newcommand{\SFRE}{$0.11^{+0.06}_{-0.04}$}
\newcommand{\massE}{$9.30^{+0.07}_{-0.10}$}
\newcommand{\beammajE}{$1.15''$}
\newcommand{\beamminE}{$0.81''$}
\newcommand{\beamPAE}{$-53.8^{\circ}$}

\newcommand{\plusminus}{±}

\usepackage{color, url}

\begin{document}

\title[Molecular gas in FRB hosts]{Constraining the Molecular Gas Content of Fast Radio Burst (FRB) Host Galaxies}

\author[0000-0002-4985-028X]{Jay S. Chittidi}
\affiliation{Department of Astrophysics and Planetary Sciences, University of Colorado, Boulder, CO 80309, USA}
\affiliation{Maria Mitchell Observatory, 4 Vestal Street, Nantucket, MA 02554, USA}

\author[0009-0001-9866-343X]{Georgia Stolle-McAllister}
\affiliation{Maria Mitchell Observatory, 4 Vestal Street, Nantucket, MA 02554, USA}
\affiliation{Museum of Science, 1 Science Park, Boston, MA 02114, USA}

\author[0000-0003-2973-0472]{Regina A. Jorgenson}
\affiliation{Maria Mitchell Observatory, 4 Vestal Street, Nantucket, MA 02554, USA}

\author[0000-0002-1883-4252]{Nicolas Tejos}
\affiliation{Instituto de F\'isica, Pontificia Universidad Cat\'olica de Valpara\'iso, Casilla 4059, Valpara\'iso, Chile}

\author[0000-0002-7738-6875]{J. Xavier Prochaska}
\affiliation{University of California - Santa Cruz, 1156 High St., Santa Cruz, CA, USA 95064}
\affiliation{Kavli Institute for the Physics and Mathematics of the Universe (Kavli IPMU), 5-1-5 Kashiwanoha, Kashiwa, 277-8583, Japan}
\affiliation{Division of Science, National Astronomical Observatory of Japan, 2-21-1 Osawa, Mitaka, Tokyo 181-8588, Japan}

\author[0000-0003-0307-9984]{Tarraneh Eftekhari}\altaffiliation{NHFP Einstein Fellow}
\affiliation{Center for Interdisciplinary Exploration and Research in Astrophysics (CIERA) and Department of Physics and Astronomy, Northwestern University, Evanston, IL 60208, USA}

\author[0000-0002-7374-935X]{Wen-fai Fong}
\affiliation{Center for Interdisciplinary Exploration and Research in Astrophysics (CIERA) and Department of Physics and Astronomy, Northwestern University, Evanston, IL 60208, USA}

\author[0000-0003-4501-8100]{Stuart D. Ryder}
\affiliation{School of Mathematical and Physical Sciences, Macquarie University, NSW 2109, Australia}
\affiliation{Astronomy, Astrophysics and Astrophotonics Research Centre, Macquarie University, Sydney, NSW 2109, Australia}

\author[0000-0002-7285-6348]{Ryan M. Shannon}
\affiliation{Centre for Astrophysics and Supercomputing, Swinburne University of Technology, Hawthorn, VIC, 3122 Australia}


\begin{abstract}
We used Bands 6 and 7 of the Atacama Large Millimeter/submillimeter Array (ALMA) in Cycles 7 and 8 to search for $\mathrm{CO}\,(3-2)$ emission from a sample of five fast radio burst (FRB) host galaxies discovered by the Commensal Real-time ASKAP Fast Transients (CRAFT) survey and the Fast and Fortunate for FRB Follow-up (F$^4$) team. These galaxies have redshifts $z \approx 0.16-0.48$, masses log$(M_{\rm star}/M_{\odot})\approx 9.30-10.4$ characteristic of field galaxies, and emission lines indicative of ongoing star formation. We detected three of the five galaxies with luminosities $L'(3-2)\approx0.2-4\times10^8\,\rm K\,km \, s^{-1}\,pc^2$ and set upper limits for the other two. Adopting standard metallicity-dependent CO-to-H$_2$ conversion factors, we estimate molecular gas masses $M_{\rm gas}\approx 0.2-3\times 10^9 \, M_{\odot}$. As a population, FRB host galaxies track the main $M_{\rm star}-M_{\rm gas}$ locus of star-forming galaxies in the present-day universe, with gas fractions of $\mu_{\rm gas}\approx0.1$ and gas depletion times $t_{\rm dep} \gtrapprox 1\,$Gyr. We employ the Kaplan-Meier estimator to compare the redshift-corrected $\mu_{\rm gas}$ and $t_{\rm dep}$ for all known FRB hosts with measurements or upper limits with those from the xCOLD GASS survey and find statistically different gas fractions. The difference is not statistically significant when we consider only the five hosts studied here with consistently determined properties, suggesting more FRB hosts with measured molecular gas masses are needed to robustly study the population. Lastly, we present a multi-wavelength analysis of one host (HG20180924B) combining high-spatial resolution imaging and integral field spectroscopy to demonstrate that future high-resolution observations will allow us to study the host galaxy environments local to the FRBs.
\end{abstract}

\section{Introduction}\label{sec:intro}

Fast Radio Bursts (FRBs) are enigmatic radio pulses that last mere milliseconds but are bright enough to be seen from cosmological distances (\citealt{lorimer2007}; see \citealt{CordesChatterjee2019} and \citealt{petroff2022} for recent reviews). The extragalactic nature of FRBs was first inferred from their large dispersion measures (DMs), whose values exceed what is expected from the Milky Way's interstellar medium (ISM) sightlines \citep[e.g.][]{lorimer2007, Spitler2016, Shannonetal2018}. Subsequent follow-up observations of well-localized ($<1\arcsec$) FRBs to individual galaxies have now confirmed their cosmological origins \citep{Tendulkar2017, Bannister2019, Ravi2019, Marcote2020, Heintz2020, Bhandari2020a, Bhardwaj2021}.

Despite significant effort from the community over the past years, the astrophysical mechanism for such bright ($\sim 10^{41}$\,erg) short pulses remains a mystery. Tens of possible progenitor scenarios have been advanced to explain FRBs \citep[e.g. see ][for a compendium]{Platts2019},\footnote{\url{https://frbtheorycat.org}} yet none is conclusive. The question of possible progenitors becomes more complicated when we consider that the population of FRBs is also divided into repeating and (apparently) non-repeating events\footnote{For convenience, we will refer to these {\it apparently} non-repeating simply as non-repeating FRBs.} \citep{Spitler2016,chime2019,Bhardwaj2021}, indicating that FRBs may be produced by more than one channel.

More than 600 FRBs have been reported (and the pace of discoveries is increasing still!) thanks to dedicated arrays and experiments such as the Canadian Hydrogen Intensity Mapping Experiment \citep[CHIME,][]{chime2014, CHIME2021}, the Australian Square Kilometer Array Pathfinder \citep[ASKAP,][]{Macquart2010, Shannonetal2018}, MeerTRAP \citep{Rajwade2022}, realfast \citep{Law2018}, and the Deep Synoptic Array \citep[DSA,][]{Kocz2019}. CHIME dominates the number of detections although with coarse ($>$arcmin) localizations, whereas the latter experiments are capable of both detecting {\it and} precisely localizing FRBs. Sub-arcsecond astrometric precision is needed for effective follow-up of galaxy hosts \citep{Eftekhari2017}, and thus far only $\sim 30$ FRB hosts have been published. Although the radio pulses themselves provide key properties (DM, rotation measure, duration, among others), multi-wavelength follow-up is likely paramount to unravel their origin(s). 

Searches for (transient) optical counterparts to FRBs have been conducted but thus far these have only yielded non-detections \citep{Hardy2017,Andreoni2020,Chen2020,Marnoch2020,Kilpatrick2021,Nunez2021, Hiramatsu2022}. Still, optical follow-up has currently been the major driver for advancing the field. For instance, spectroscopic redshifts of FRB galaxy hosts has constrained
their possible progenitors (e.g., \citealt{Bhandari2020b}), and enabled their use as cosmological probes \citep{Macquart2020,Hagstotz2022,James2022,Ryder2022}. 
The study of the FRB host galaxy population can also constrain theoretical models by assessing possible correlations between FRBs, stellar populations, gas content, metallicity, kinematics and/or locations within their galaxies \citep{Mannings2020,Chittidi2021}.
Pioneering work on FRB host galaxy demographics has shown diverse properties in stellar masses and star formation rates \citep{Heintz2020,Bhandari2020a,Gordon2023}, spanning virtually the full range expected for field galaxies at their corresponding redshifts. In particular, FRB hosts have stellar masses of M$_{*} \approx 10^8-10^{11}$\,M$_{\odot}$, SFRs $\approx $ 0.05 -- 36 M$_{\odot}$ yr$^{-1}$ \citep{Gordon2023}, and a range of morphologies spanning dwarfs, spirals, and ellipticals \citep{Sharma2023}. Several FRBs appear to be coincident with spiral arm sub-structure of their hosts, providing a potential link to recent star formation \citep{Mannings2020}. In contrast, some FRBs also appear within galaxies dominated by older stellar populations \citep{Bannister2019, Ravi2019,Sharma2023}. 

In the local Universe, the association of the Milky Way magnetar SGR~1935+2514 with an FRB-like burst \citep{CHIME2020magnetar} provides strong evidence that FRBs could in principle be powered by magnetars, providing further support to the association with young stellar populations. Conversely, the localization of a FRB originating from within a globular cluster in Messier 81 \citep{Bhardwaj2021,Kirsten2022} points towards a much older stellar population.

Follow-up at other wavelengths is required to complement 
efforts in the optical, especially for assessing the connection between star formation and FRB progenitors. 
For instance, mapping the cold gas content and kinematics could provide new clues, as violent processes that give rise to star formation often leave indelible imprints upon the neutral gas, e.g. in the form of significant disturbances in the velocity field \citep{Arabsalmani2019a,Roychowdhury2019}. Indeed, optical imaging of FRB hosts has suggested that some events may occur in perturbed star-forming environments \citep[e.g.][]{Marcote2020,Bhandari2020b}. Mapping of the 21-cm neutral atomic hydrogen gas in a FRB host galaxy presented by \cite{Kaur2022} suggests the galaxy has recently undergone a minor merger that may have caused a burst of star formation giving rise to the FRB progenitor. On the other hand, if FRB hosts are deficient in molecular gas, one may then conclude that channels of production from older stellar populations are preferred instead. Thus, probing the physical condition of the cold molecular gas (which fuels star formation) via sub-mm observations seems a timely endeavor.

Few searches for molecular gas in FRB host galaxies have been completed, and these prior studies have had limited success. For instance, \cite{Bower2018} placed upper limits on the CO luminosity for the host galaxy of FRB\,20121102A
(HG20121102A; throughout the paper we designate the
name of host galaxies with HG), 
resulting in an upper limit on the H$_2$ mass of $2.5 \times  10^8 \ {\rm M}_{\odot}$. More recently, \cite{Hatsukade2022} presented a sample containing one new detection of molecular gas in the host galaxy of FRB\,20180924B,\footnote{We note that this detection was first presented on 3 August 2022 at the IAU Symposium 369 by our collaboration, e.g. \url{https://www.youtube.com/watch?v=qz92y4AtpDo}} and two upper limits for FRB\,20190102C, and FRB\,20190711A, which they compare with two local FRBs (20200120E and 20200428A).

In this paper, we expand on these previous works by quantifying the molecular gas mass ($M_{\rm gas}$), gas fraction ($\mu_{\rm gas}$), and depletion timescale ($t_{\rm dep}$) for three hosts with new CO observations. Combined with existing observations, this work represents the largest compilation of CO observations in FRB hosts to date. Our results are used to offer the first general census of the molecular gas content of FRB hosts and to establish their connection to other relevant physical parameters, such as the SFR and stellar mass, needed for determining the ISM conditions, and thus pinpointing the stellar progenitors that could give rise to FRBs.  

Our paper is structured as follows. In Section~\ref{sec:sample} we present our sample selection. In Section~\ref{sec:ALMAobs} we present our ALMA observations and data reduction procedure. Section~\ref{sec:ALMAanalysis} presents our main analysis and results, which we discuss in Section~\ref{sec:discussion}. A summary and conclusions are given in Section~\ref{sec:summary}. Throughout this paper, we use the Planck15 cosmology \citep[$H_0$=67.74 km s$^{-1}$ Mpc$^{-1}$, $\Omega_{\rm M}$ = 0.3075,][]{Planck15} implemented in the \code{astropy.cosmology} package.

\section{The Sample}
\label{sec:sample}
\begin{table*}[t]
\centering
\textbf{FRB and Host Galaxy Coordinates} \\
\begin{tabular}{cccccc}
\hline
\hline
FRB & FRB RA & FRB Dec & Host RA & Host Dec\\
& J2000 & J2000 & J2000 & J2000  \\
\hline
20180924B & \RAA\ \plusminus\ \RAerrA & \DecA\ \plusminus\ \DecerrA & \hostRAA & \hostDecA  \\
20181112A & \RAB\ \plusminus\ \RAerrB & \DecB\ \plusminus\ \DecerrB & \hostRAB & \hostDecB  \\
20190102C & \RAC\ \plusminus\ \RAerrC & \DecC\ \plusminus\ \DecerrC & \hostRAC & \hostDecC  \\
20190714A & \RAD\ \plusminus\ \RAerrD & \DecD\ \plusminus\ \DecerrD & \hostRAD & \hostDecD  \\
20200430A & \RAE\ \plusminus\ \RAerrE & \DecE\ \plusminus\ \DecerrE & \hostRAE & \hostDecE  \\
\hline
\end{tabular}
\caption{Positions of FRBs and their host galaxies. The FRB coordinates and total uncertainties (systematic and statistical summed in quadrature) are from \cite{Day2021}. The host galaxy coordinates and redshifts are from the public FRB repository \citep{frbrepo}.}
\label{tab:coords}
\end{table*}
Our sample consists of five FRB host galaxies observed with ALMA during Cycles 7 and 8.  The sample host galaxies were taken from well-localized FRB detections, all of which were detected by the Commensal Real-time ASKAP Fast Transients \citep[CRAFT,][]{Macquart2010} survey, using ASKAP \citep{Mcconnelletal2016} between 2018 and 2020. A summary of the five FRBs and their host galaxy positions is given in Table \ref{tab:coords}, while the host galaxy properties are provided in Table \ref{tab:sample}. 

\cite{Bannister2019} presented the first detection and localization of FRB\,20180924B and used it as a probe of baryonic content in the IGM; as part of that analysis they also reported on properties of the host galaxy (HG). \cite{Prochaska2019} reported the detection and localization of FRB\,20181112A and used it to measure properties of a foreground galaxy, but also reported on some properties of the host galaxy. \cite{Macquart2020} presented the detection and localization of FRB\,20190102C, and used a selection of galaxies (including HG20180924B, and HG20181112A) to estimate the contribution of host galaxies to their observed FRB dispersion measures in order to measure the baryonic content of the IGM. These three galaxies were also included in \cite{Bhandari2020a}, which investigated the global properties of the first four host galaxies associated with FRBs localized by ASKAP. In \cite{Heintz2020}, the authors presented the localizations of FRB\,20190714A and FRB\,20200430A and analyzed a larger sample of galaxies, including all five examined here, to compare characteristics of FRB host galaxies. \cite{Mannings2020} gave a high resolution view of HG20180924B, HG20190102C, and HG20190714A along with several other FRB host galaxies using HST to examine the position of the FRBs within their host galaxies. \cite{Gordon2023} probed the star formation history of a large sample of FRB host galaxies, including the five presented here, to better understand the implications for FRB progenitors. All of the FRB localizations were updated in \cite{Day2021} and the host galaxy coordinates were retrieved from the public FRB repository \citep{frbrepo}.

\begin{table}[h]
\centering
\textbf{FRB Host Galaxy Sample Properties} \\
\begin{tabular}{ccccc}
\hline
\hline
Host & z & log(Z$_{gas}$/Z$_{\odot}$) & log(M$_{*}$/M$_{\odot}$) & SFR\\
&  &  &  & M$_{\odot}$ yr$^{-1}$  \\
\hline
20180924B & \zA & \metA & \massA & \SFRA \\
20181112A & \zB & \metB & \massB & \SFRB \\
20190102C & \zC & \metC & \massC & \SFRC \\
20190714A & \zD & \metD & \massD & \SFRD \\
20200430A & \zE & \metE & \massE & \SFRE \\
\hline
\end{tabular}
\caption{Summary of sample properties. The host galaxy redshifts are from the public FRB repository \citep{frbrepo}. 
The log(Z$_{gas}$/Z$_{\odot}$) is the gas-phase metallicity; log(M$_{*}$/M$_{\odot}$) is the stellar mass of the galaxy; and SFR is the star formation rate integrated over $0-100$ Myr, all as reported
in \cite{Gordon2023}.}
\label{tab:sample}
\end{table}

\section{ALMA Observations and Data Reduction}\label{sec:ALMAobs}

We obtained ALMA observations of five FRB host galaxies taken during ALMA Cycles 7 and 8. A brief summary of the ALMA observations and the naturally-weighted \texttt{CLEAN} beam sizes are presented in Table~\ref{tab:almaobs}. Observations for the three galaxies targeted the CO $J=3-2$ (hereafter $\co$) transition, the strongest of the low-$J$ transitions expected to be redshifted into the ALMA bands (namely Bands 6 and 7 for this line). As the goal of the observations was to detect (or put a limit on) the total line flux for each galaxy, the angular resolutions were set to the approximate sizes of the galaxies from optical data in order to avoid resolving out the emission, but the three detections appear modestly resolved. The spectral setup used 1875 MHz bandwidth with a resolution of 7812.5 kHz (after spectral averaging to reduce the data rate), which ensured that $>$15 resolution elements covered the expected 150~km\,s$^{-1}$ FWHM linewidth. However, we further averaged the data during imaging to a resolution of 39.063 MHz in order to confidently measure the integrated CO fluxes. The correlator was set to time-division mode (TDM) with 31.250 MHz resolution for continuum detections near the CO line for each galaxy, but none were detected. 

The data were processed through the standard ALMA data reduction pipeline and then imaged using the Common Astronomy Software Applications \citep[\texttt{CASA} v6.5.3,][]{CASA} \texttt{TCLEAN} task to produce image cubes. For each target, we produced an initial dirty image cube from which we measured the RMS in 10 regions away from the expected position of the galaxy. Then, the cubes were \texttt{CLEAN}ed to the RMS level and primary beam-corrected.

$\co$ emission from the host galaxies HG20180924B, HG20190714A, and HG20200430A, was detected above 3$-\sigma$ significance with linewidths of FWHM\,$\approx$\,320 km s$^{-1}$. We generated moment-0 (velocity-integrated) maps of the image cubes by integrating over the -160 $\kms$ to +160 $\kms$ using the \texttt{astropy}-affiliated \texttt{Python} package \texttt{spectral-cube}\footnote{\url{https://spectral-cube.readthedocs.io}} \citep{spectralcube}. Then, we extracted 1-D spectra using a region the size of the synthesized beam centered on the brightest pixel in the moment-0 maps.  HG20181112A and HG20190102C were not detected, and so we placed $3-\sigma$ upper limits on the emission using spectra that were extracted in the same manner as the detections but from the expected position of the host galaxies.

A bright sub-mm galaxy (SMG) was detected $6''$ west of the expected position of HG20181112A, distinct from a previously detected bright foreground galaxy $\approx5''$ north of the host found in $g$-band FORS2 imaging by \citet{Prochaska2019}, but the SMG found here is absent in that data. Likewise, the northern foreground galaxy is not detected in the ALMA data. If the SMG is foreground to HG20181112A, its baryonic halo could contribute to the detected dispersion measure (DM) of the FRB. However, bright SMGs in Band 6 tend to correspond to higher redshift galaxies \citep{Chen2023}, and since we detect the SMG in continuum emission rather than line emission, we cannot constrain the redshift of the galaxy well, and we do not study the source further.

\begin{figure*}[h]
\centering
\includegraphics[width=0.52\textwidth]{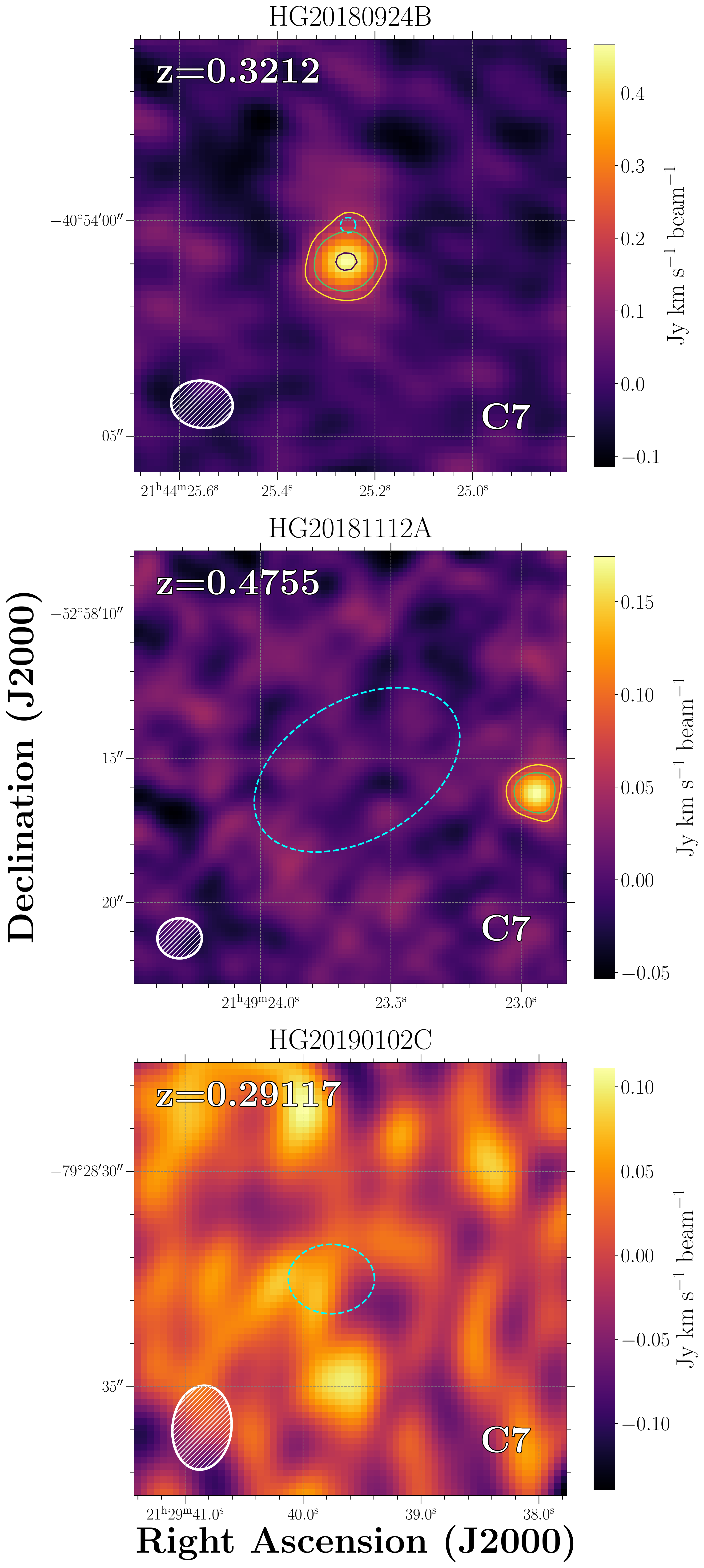}
\includegraphics[width=0.415\textwidth]{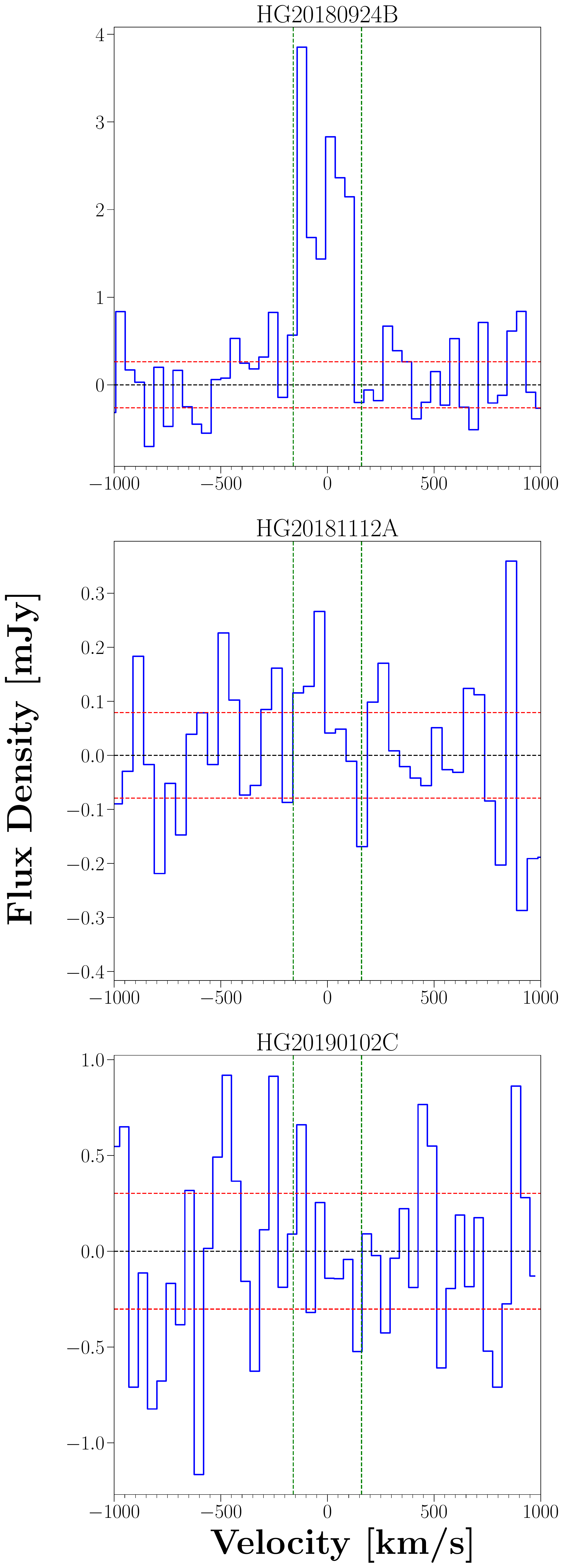}
\caption{{\it Left column}: Moment-0 (velocity-integrated) map of $\co$ emission from FRB host galaxies HG20180942B, HG20181112AA, and HG20190102C observed during Cycle 7 (C7). The average synthesized beam across the integrated channels is plotted in the lower left. The FRB localizations are indicated by the cyan dotted ellipses.  For the detection in HG20180924B, contour levels are 3, 5, and 10 times the RMS measured in the moment-0 image. {\it Right column:} CO spectra extracted from a region the size of the synthesized beam centered on the brightest pixel in the moment-0 image, or the host galaxy coordinates for the non-detections. The green vertical dashed lines indicate the $\pm$160$\kms$ range over which the line was integrated, while red horizontal dashed lines indicate the $\pm$1-$\sigma$ errors on the spectrum as measured in the $500 - 800 \kms$ channels.}
\label{fig:alma_c7}
\end{figure*}

\begin{figure*}[h]
\centering
\includegraphics[width=0.52\textwidth]{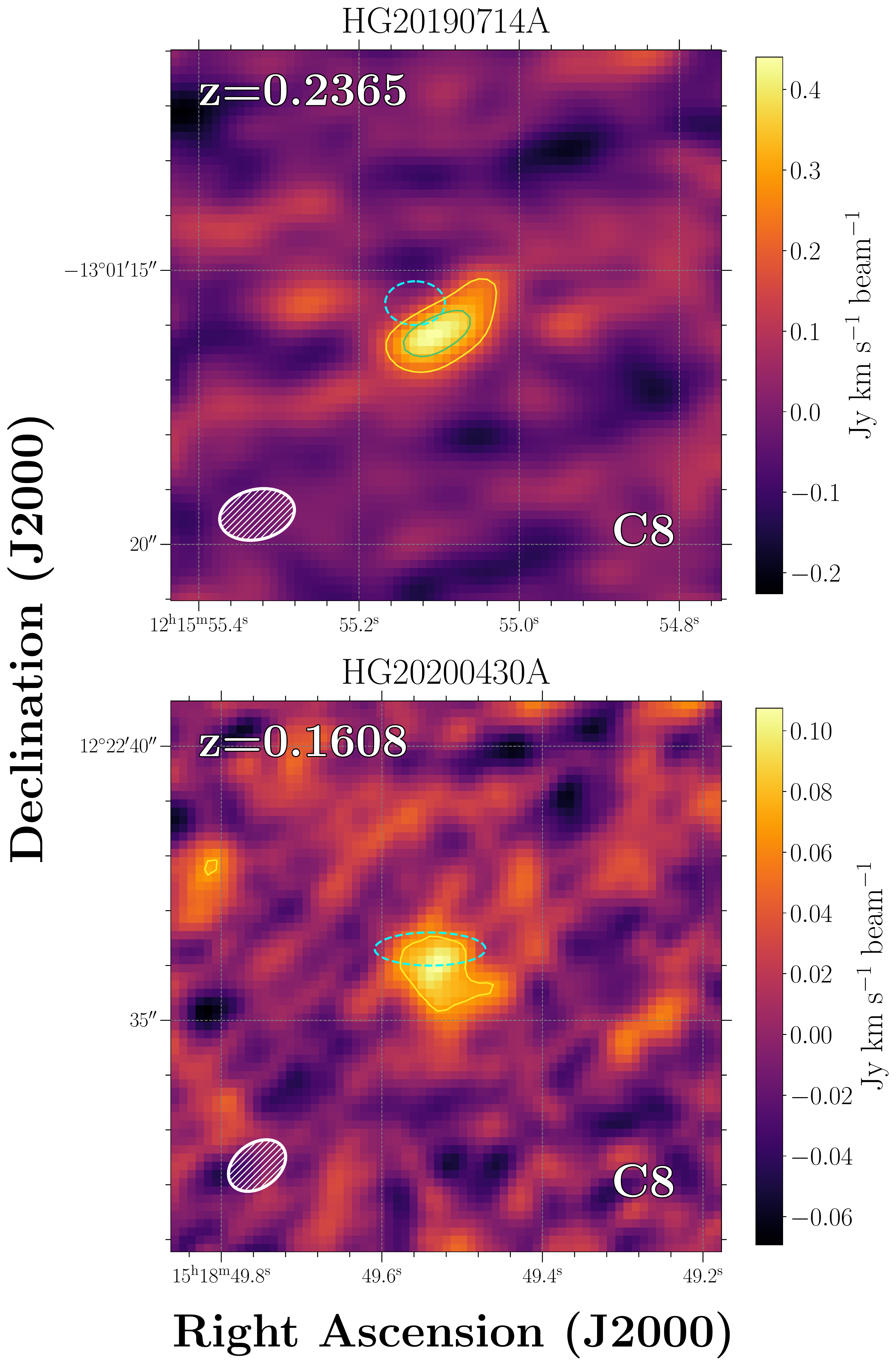}
\includegraphics[width=0.425\textwidth]{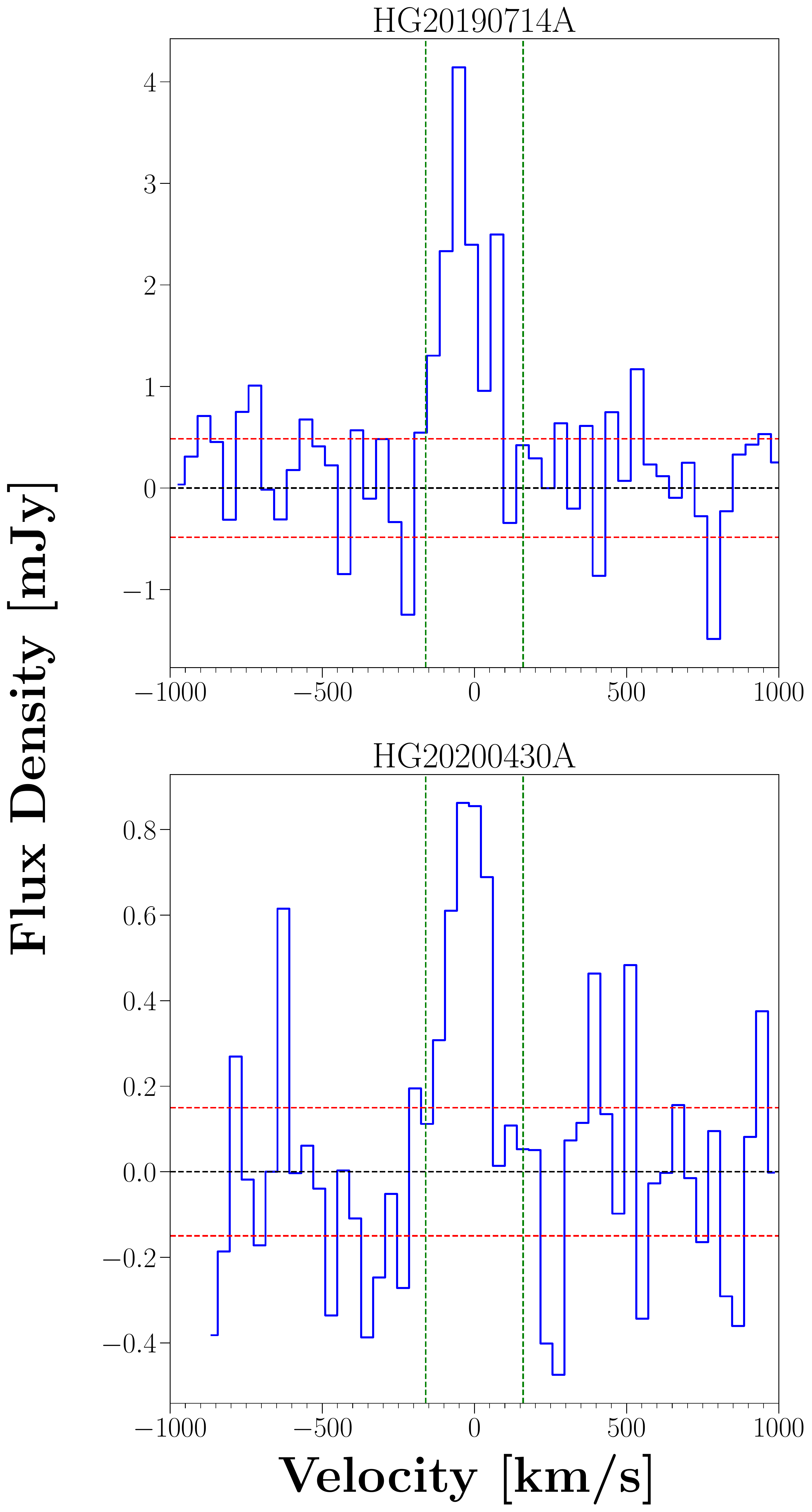}
\caption{Same as Figure \ref{fig:alma_c7}, but for the Cycle 8 (C8) targets HG20190714A and HG20200430A.}
\label{fig:alma_c8}
\end{figure*}

In Figures \ref{fig:alma_c7} and ~\ref{fig:alma_c8} we present the velocity-integrated intensity maps of the $\co$ emission and the extracted spectra of the host galaxies observed in cycles 7 and 8, respectively. The characteristic RMS of the spectra were calculated from channels offset from the predicted spectral lines in the range $500-800 \kms$. The central frequencies of the line, the RMS of the spectra, velocity channel widths, and the resultant velocity-integrated fluxes are listed in Table \ref{tab:alma_results}.

\begin{table}[h]
\centering
\textbf{ALMA Observations} \\
\begin{tabular}{cccccc}
\hline
\hline
Host & Obs Date & Config. &\multicolumn{3}{c}{Beam Size}  \\
 & UTC & & major & minor & PA \\ \hline
20180924B & \obsdateA & C43-1 & \beammajA & \beamminA & \beamPAA  \\
& 2019-11-27 & &&&\\
20181112A & \obsdateB & C43-1 & \beammajB & \beamminB & \beamPAB  \\
& 2019-12-20 & &&& \\
& 2022-03-06 & &&& \\
20190102C & \obsdateC & C43-1 & \beammajC & \beamminC & \beamPAC  \\
& 2019-12-03 & &&& \\
20190714A & \obsdateD & C43-1 & \beammajD & \beamminD & \beamPAD \\
20200430A & \obsdateE & C43-2 & \beammajE & \beamminE & \beamPAE \\
& 2022-04-19 & &&&\\
\hline
\end{tabular}
\caption{Summary of observations. The beam size is the ellipse of the average synthesized beam across the integrated channels, with the major and minor axes measured in arcseconds and the position angle (PA) in degrees east of north. Cycle refers to the ALMA observing cycle.}
\label{tab:almaobs}
\end{table}


\section{ALMA Analysis and Results}\label{sec:ALMAanalysis}

We estimate the molecular gas masses and upper limits for the sample using standard procedures.  First, we calculate the CO line luminosity, $L'_{\rm CO}$ in units of K\;km\;s$^{-1}$\;pc$^2$, using equation~3 from \cite{solomon2005}, 

\begin{equation}\label{line_lum}
 L'_{\rm CO} = 3.25 \times 10^7 S_{\rm CO} \Delta v \nu ^{-2}_{\rm obs} D^2 _L (1+z)^{-3}   
\end{equation}

\noindent where $S_{\rm CO} \Delta v$ is the velocity integrated flux in units of Jy $\kms$, $\nu_{obs}$ is the observed frequency of the $\co$ line in GHz, and D$_L$ is the luminosity distance in Mpc. The linewidths for detected galaxies in this sample are all consistent with $320 \kms$, so we use this value for calculating $S_{\rm CO} \Delta v$. Following equation 4 from \cite{solomon2005}, the molecular gas mass, $M_{\rm gas}$, is defined as 

\begin{equation}
    M_{\rm gas}=\alpha L'_{\rm CO} 
\end{equation}

\noindent where $\alpha$ is the CO-to-H$_2$ conversion factor.  

As $\alpha$ is dependent upon the gas metallicity, we follow the procedure outlined in \cite{genzel2015}, equations 6, 7, and 8, to determine $\alpha$ for each host galaxy using

\begin{equation}
    \alpha _{03} = \alpha _{MW} \times \chi (Z) \times R_{13}
\end{equation}

\noindent where we take the Milky Way value of $\alpha$ to be $\alpha _{MW}$=4.3 M$_{\odot}$(K $\kms \rm pc^2)^{-1}$ \citep{Bolatto2013}. $\chi (Z)$ is a fitting function dependent on the gas metallicity, for which we use the metallicity measured by \cite{Gordon2023} and listed here in Table \ref{tab:sample} for reference. Following \cite{genzel2015}, we take the mean of two prescriptions for $\chi(Z)$ except for HG20190102C, which has a $\mathrm{log}(Z/Z_{\odot})<-0.27$, placing it in the metallicity regime for which the two prescriptions differ significantly. Since this host was not detected, we select the $\chi(Z)$ relationship in Equation 7 of \citet{genzel2015} which results in a higher, and thus more conservative, upper molecular gas mass limit. R$_{1J}$ is the ratio of the $\ensuremath{(1-0)}$ to the $\ensuremath{(3-2)}$ CO line luminosity, R$_{13} ={L}_{\mathrm{CO}(1\mbox{--}0)}^{{\prime} }/{L}_{\mathrm{CO}(3\mbox{--}2)}^{{\prime} }$, for which we use 1/0.55 ~\cite{Lamperti2020}, as a typical value for local star-forming galaxies.

$\co$ emission from HG20181112A and HG20190102C was not detected, but we placed 3-$\sigma$ molecular gas upper limits of 1.29$\times 10^9$ $M_{\odot}$ and 4.34$\times 10^9$ $M_{\odot}$ using the same method above, where $\sigma$ is the RMS measured in the spectra as described in Section \ref{sec:ALMAobs}.

\begin{figure*}[h]
    \centering
    \includegraphics[width=0.48\textwidth]{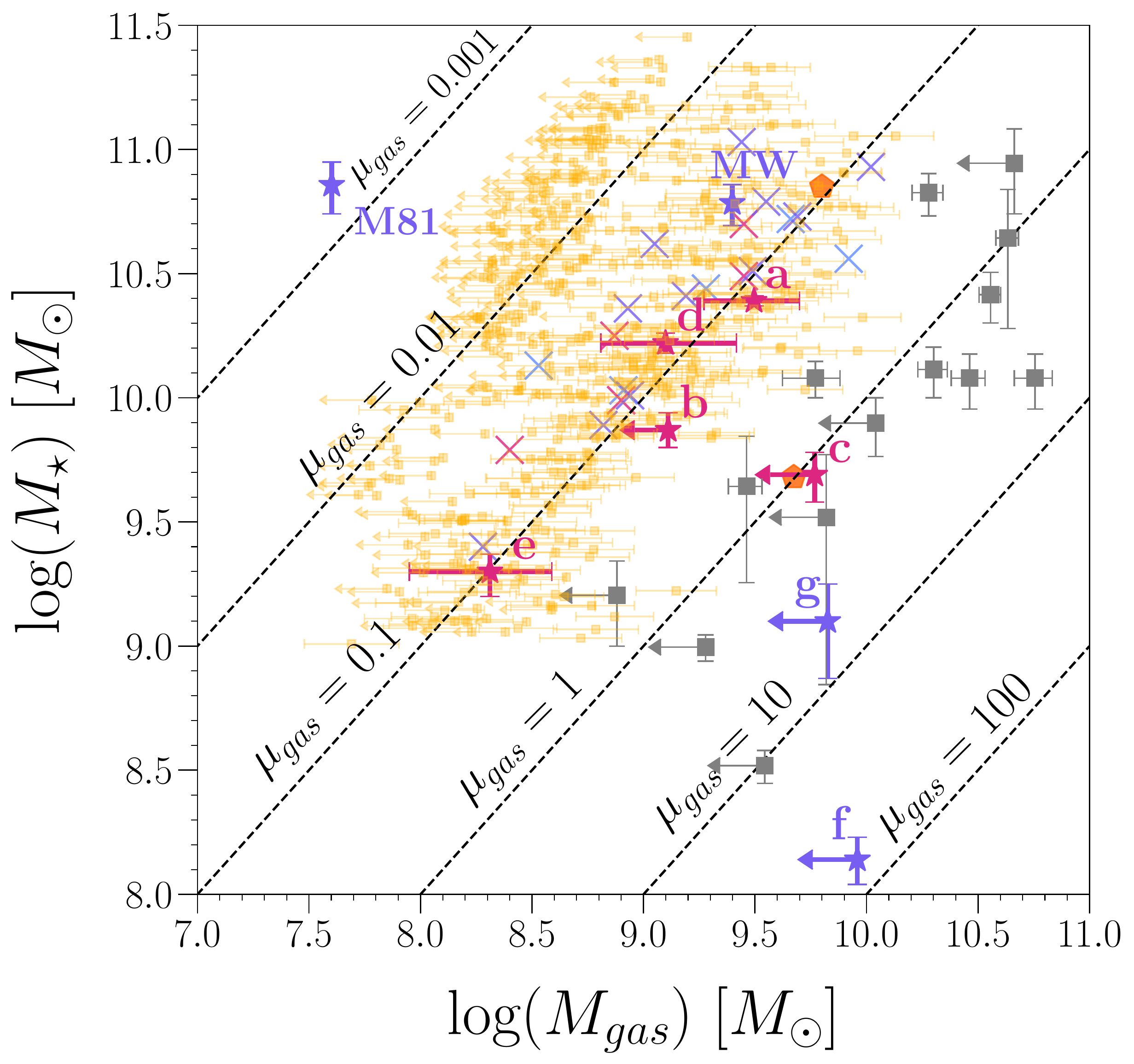}
    \includegraphics[width=0.49\textwidth]{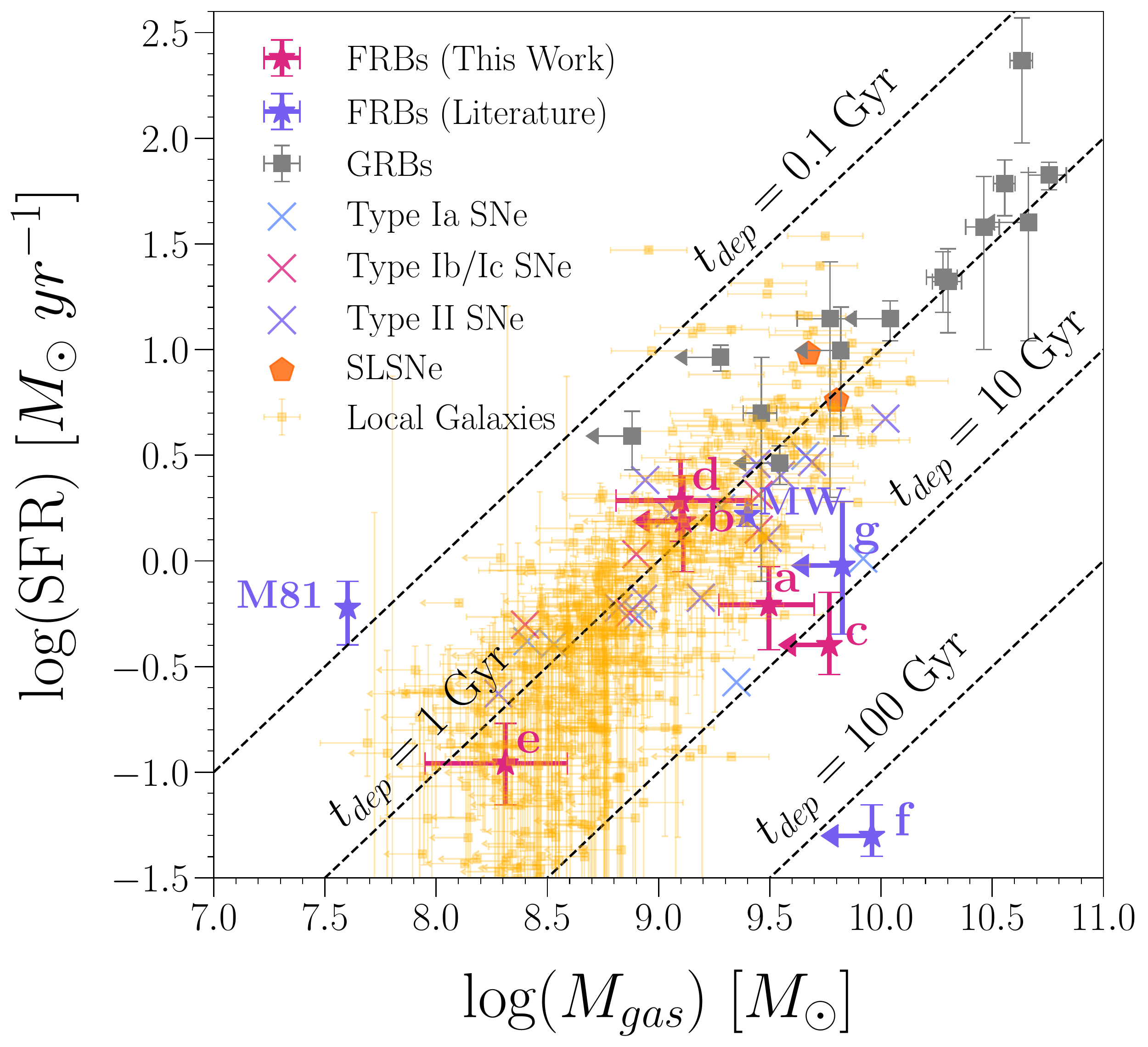}
    \caption{ {\it Left:} Stellar mass vs. molecular gas mass for the five FRB hosts observed with ALMA in Cycle 7 and Cycle 8 (red stars, arrows represent 3-$\sigma$ upper limits), compared to long GRB hosts \citep[gray squares,][]{Hatsukade2020_grb}, SNe populations \citep[colored-x's,][]{Galbany2017}, SLSNe \citep[orange pentagons,][]{Arabsalmani2019b,Hatsukade2020_slsn}, and local galaxies from the IRAM Survey \citep[yellow squares,][]{saintonge2017}. The population of local galaxies at $\mu_{\rm gas}$=0.01 are upper limits from \citet{saintonge2017}, and are used for the statistical analysis in Section \ref{sec:disc_stats}. Host galaxies are labeled as follows: a) HG20180924B, b) HG20181112A, c) HG20190102C, d) HG20190714A, e) HG20200430A.  Overplotted from the literature (purple stars, arrows represent 3-$\sigma$ upper limits) are host galaxies: f) 20121102A, g) 20190711A, M81)  20200120E, and MW) 20200428A ~\citep{Hatsukade2022}.  Dashed lines denote lines of constant gas fraction, $\mu_{\rm gas}$. {\it Right:} SFR vs. molecular gas mass for the same galaxies; dashed lines represent constant depletion timescales, $t_{\rm dep}$. At face value, FRB hosts with detections are consistent with the distribution of local field galaxies, with $t_{\rm dep} \lesssim 5$~Gyr and $\mu_{\rm gas}\approx0.1$.}
    \label{fig:mgas}
\end{figure*}

In Figure~\ref{fig:mgas}, we present the stellar mass versus molecular gas mass, and star formation rate (SFR) versus molecular gas mass relations for the sample.  These plots trace the gas fraction, $\mu_{\rm gas}=M_{\rm gas}/M_{\star}$, and gas depletion time, $t_{\rm dep}=M_{\rm gas}/$SFR.  For comparison, we overplot values from host galaxies of long-duration GRBs (LGRBs) ($0.3 <z<2$) studied with ALMA \citep{Hatsukade2020_grb} and local galaxies ($0.01 < z < 0.05$) from the IRAM Survey \citep{saintonge2017}. We also plot values for supernovae host galaxies from \citet{Galbany2017} and super-luminous supernovae hosts from \citet{Arabsalmani2019b} and \citet{Hatsukade2020_slsn}. Additionally, we plot FRB hosts taken from the literature including the host galaxy of FRB\,20121102A, which was reported to have a molecular gas upper limit of $< 9.1 \times 10^9$ M$_{\odot}$ by \cite{Hatsukade2022}.  For this source, we use the SFR and stellar mass values reported by \cite{Gordon2023}. 
Three additional FRB host sources were taken from the literature including the M$_{gas}$ upper limit on HG20190711A,
and the values derived for the M81 globular cluster and Milky Way (MW) FRBs, 20200120E and 20200428A, respectively \citep{Hatsukade2022}.

\begin{table*}[h]
\centering
\textbf{ALMA Results}
\begin{tabular}{cccccccccc}
\hline
\hline
FRB Host Galaxy & $\nu$(3-2) & RMS & $\Delta v$ & $S_{\rm CO} \Delta v$ & $L'(3-2)$ & $M_{\rm gas}$ & $t_{\rm dep}$ & $\mu_{\rm gas}$ \\
& GHz & mJy & km s$^{-1}$ & Jy km s$^{-1}$  & 10$^8$ K km s$^{-1}$ pc$^2$ & 10$^9$ $M_{\odot}$ & Gyr &  \\
\hline
20180924B & 261.728 & 0.26 & 45 & 0.69 $\pm$ 0.094 & 4.26 $\pm$ 0.58 & $3.13^{+1.87}_{-1.27}$ & $5.05^{+8.12}_{-3.07}$ & $0.13^{+0.09}_{-0.05}$\\
20181112A & 234.358 & 0.08 & 50 & $<$ 0.084  & $<$ 1.17 & $<$ 1.29 & $<0.84$& $<0.17$\\
20190102C & 267.816 & 0.30 & 44 & $<$ 0.33  & $<$ 1.66 & $<$ 5.86 & $<14.6$ & $<1.20$\\
20190714A & 279.657 & 0.48 & 42 & 0.6 $\pm$ 0.068 & 1.97 $\pm$ 0.22 & $1.26^{+1.36}_{-0.62}$ & $0.65^{+1.46}_{-0.44}$& $0.08^{+0.10}_{-0.04}$\\
20200430A & 297.894 & 0.15 & 40 & 0.14 $\pm$ 0.053 & 0.21 $\pm$ 0.08 & $0.20^{+0.18}_{-0.12}$ & $1.86^{+3.69}_{-1.34}$ & $0.10^{+0.13}_{-0.06}$\\
\hline
\end{tabular}
\caption{Measured and inferred properties from the extracted $\co$ spectra from the ALMA observations. Values for HG20181112A and HG20190102C are $3-\sigma$ upper limits.}
\label{tab:alma_results}
\end{table*}


\section{Discussion}\label{sec:discussion}

\subsection{Comparing FRB Hosts with a sample of local galaxies} \label{sec:disc_stats}

Given the low number of CO detections, a cursory look at Figure \ref{fig:mgas} suggests that the C7/C8 sample of this work exhibits gas fractions ($\mu_{\rm gas}\approx0.1$) and depletion times ($t_{\rm dep} \lesssim 5$~Gyr) that are comparable to local galaxies. However we aim to more robustly test if the two populations are similar by following the statistical procedure laid out in \citet{Hatsukade2022} to apply the Kaplan-Meier estimator \citep{KaplanMeier1958}, which gives weight to the four non-detected hosts. Since the \citet{saintonge2017} sample of local galaxies are at much smaller redshifts than the FRB hosts, we compute the redshift independent gas fractions and depletion times for the two samples by following the fits in Table 3 of \citet{Tacconi2018} (see rows ``CO(error w.)" for $t_{\rm dep}$ and ``\textbf{Best}, $\beta=0$" that uses Main Sequence fits from \citet{Speagle2014} for $\mu_{\rm dep}$) that account for the redshift, stellar mass, and the offset relative to a fit to the Main Sequence line of galaxies.

We fit the survival functions of the corrected gas fractions and depletion times for the local galaxies and FRB hosts with the Kaplan-Meier estimator using the \texttt{Python} package \texttt{lifelines} \citep{Davidson-Pilon2019} and compare them with the respective functions for the \citet{saintonge2017} sample. The resulting functions are similar within the confidence interval, so we perform the log-rank test to statistically analyze the difference. The resulting $p$-values for $\mu_{\rm gas}$ and $t_{\rm dep}$ are 0.002 and 0.60, suggesting that the gas fractions are from statistically different distributions than local galaxies in the \citet{saintonge2017} sample, but the depletion times are not. \citet{Hatsukade2022} reported a similar possibly significant difference in $\mu_{\rm gas}$ but found that the result is not robust against outliers like M81, which has the smallest $\mu_{\rm gas}$ in the sample. We find that the $p$-value \textit{decreases} to the 99.8\%\ confidence level when removing M81 from the sample, suggesting that the majority of FRB host environments may come from galaxies with higher $\mu_{\rm gas}$ than local galaxies. 
But when we consider only the sample of five observations presented in this work (for which our values are self-consistently determined), the difference becomes insignificant ($p$=0.10). Lastly, the survival functions of $\mu_{\rm gas}$ for the FRB hosts and the local galaxy sample cross, which can lead to inaccurate results from the log-rank test; 
therefore, we are cautious against 
over-interpreting this result.

Given the emerging evidence that FRBs may occur in different galactic environments and from different progenitors, it is not surprising that this overarching sample of hosts may have properties distinct from local galaxies. The most unique FRB environments, FRB\,20200120E in a globular cluster of M81 and FRB\,20121102A in a dwarf galaxy, are from repeaters, while the detections here are from apparent non-repeaters. If a population of FRBs indeed occur in old stellar environments like FRB\,20200120E, constraints on their gas fraction from CO observations could help identify them. The significance of the statistical result discussed above can be greatly improved with more CO detections of FRB hosts, and new distinctions may emerge when enough measurements are made for repeaters and non-repeaters. There are at least four more well-localized FRBs whose host galaxy CO lines are redshifted into the ALMA bands, and have yet to be observed as of the end of C8.

\subsection{Multiwavelength, high resolution follow-up of FRB host galaxies}

Detailed observations of FRB environments play a key role in understanding the stellar populations that lead to FRB source production.  Higher resolution, multi-wavelength observations will thoroughly measure the global 
properties of the host galaxies, as well as the physical properties in the immediate vicinity of the well-localized FRBs.

\begin{figure*}[h]
\centering
\includegraphics[width=1.0\textwidth]{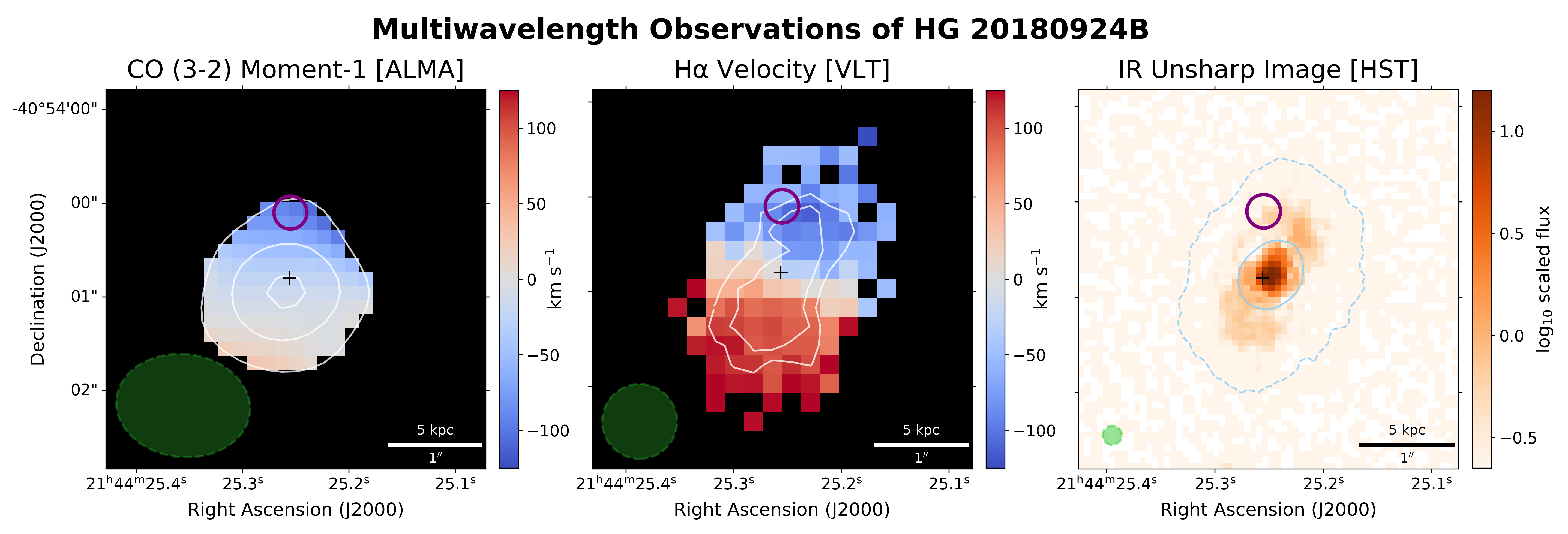}
\caption{Multiwavelength comparison of HG20180924B. Across all three images the spatial scale is the same, the black plus sign marks the center of the galaxy, and the purple ellipse marks the FRB position and its uncertainty. The shaded green ellipses are the beam size, seeing, and PSF of the observations, left to right, respectively. The image on the left is the moment-1 map of $\co$ emission from ALMA; the center image shows H$\alpha$ velocity detected using the MUSE instrument at the VLT; and the image on the right is a visually-enhanced IR HST image. The details and other features are explained in the text.}
\label{fig:20180924compare}
\end{figure*}

In addition to using ALMA observations to study the molecular content of the host galaxies, we can also compare them with other follow-up observations in optical and infrared wavelengths to produce a more complete picture.  For example, in the case of HG20180924B, we use optical integral field unit (IFU) observations from VLT/MUSE \citep{Bannister2019} and described in ~\cite{Simha2021}, and infrared observations from HST \citep{Mannings2020} to pair with the ALMA observation and gain additional information about the galaxy, as shown in Figure~\ref{fig:20180924compare}. The ALMA image (left panel) shows the moment-1 map of $\co$ emission generated using \texttt{spectral-cube} with contours representing 3, 6, and 9 times the RMS of the velocity-integrated intensity. 

With high signal-to-noise ratio (SNR) optical VLT/MUSE \citep{MUSE} IFU observations we can map the emission flux and kinematics of the ionized gas in the galaxy. In the middle panel of Figure~\ref{fig:20180924compare} we show the velocity map of H${\alpha}$, the strongest line in the observation, which we calculated by fitting a Gaussian to the emission line in every spaxel and setting the zero velocity to be at the galaxy's redshift $z=0.3212$. The contours on the plot mark H${\alpha}$ flux SNR = 15 and 10, and we applied a mask to remove all spaxels with H${\alpha}$ flux SNR$<$4. The shaded green circle in the lower left corner shows the seeing of the observation, which we determined to be $0.789\arcsec$\plusminus$0.003\arcsec$ by examining the PSFs of three stars in the wider field of view. 

HST offers high resolution images, and \cite{Mannings2020} presents images of many FRB host galaxies in both UV and IR. We used the HST IR image of HG20180924B and applied {\it unsharp} processing to visually enhance the image. This is a linear image processing technique, which we calculated as follows:

\begin{equation}
 u = g+C*(g-f(g,k))
\end{equation}

\noindent where $g$ is the original image data, and $f()$ is the smoothing convolution function with Gaussian kernel, $k$, which is built using the point spread function (PSF) of the HST observation. We chose a scaling constant $C=10$ after trying several different constants because we found that over-enhancing the image brought out features more clearly. We then plotted the {\it unsharp} image in logarithmic intensity and in this case, revealing faint spiral arms in the galaxy. The contours on the image show the log$_{10}$ of the original data without the {\it unsharp} filter and the shaded green circle in the bottom left shows the FWHM of the PSF of the observation, which was approximately $0.2 \arcsec$ \citep{Mannings2020}.

Since the uncertainty on the localization of this FRB is very small ($ \lesssim 0.2\arcsec $), we can compare the physical properties at the precise location of the FRB with those of the rest of the galaxy, making the
multi-wavelength observations even more valuable.  From the HST image in Figure~\ref{fig:20180924compare} (right panel), it is clear that the FRB localization falls on one of the spiral arms of the galaxy, as discussed previously in ~\cite{Mannings2020}.  Generally, the velocity maps of the molecular and ionized gas traced by CO and H$\alpha$ (left and middle panels of Figure~\ref{fig:20180924compare}, respectively) appear to show similar kinematic structures.  The velocity and velocity dispersion of H$\alpha$ emission is consistent with disk rotation, and detailed inspection does not reveal any additional disturbed kinematic signature at the location of the FRB. Using a higher velocity resolution of 9$\kms$, \cite{Hsu2023} report a positional offset between the two velocity components of CO emission and suggest this is indicative of a disturbed kinematic structure of molecular gas in the host galaxy.  

High angular resolution integral field unit studies of spiral galaxy hosts of FRBs like those done in \citet{Chittidi2021} and \citet{Tendulkar2021} helped constrain the local SFRs at the FRB positions, and provided unique constraints on the possible age of neutron star progenitors assuming that their formation was triggered by the spiral density wave of the galaxy. A future high-resolution ALMA CO study of HG20180924B could resolve the spiral structure seen in the HST image and is hinted at in the ALMA data by the minor contour elongation towards the northern direction in the moment-0 image.

\section{Summary and Conclusions}
\label{sec:summary}

In this letter, we analyzed ALMA Bands 6 and 7 observations of $\co$ emission from five FRB host galaxies, three of which are studied for the first time here.  We present the detection of molecular gas in two new FRB host galaxies, tripling the existing number of detections. We now summarize our findings:

\begin{enumerate}
    \item The detected hosts HG20180924B, HG20190714A, and HG20200430A have $M_{\rm gas}=0.2-3.13\times10^9M_{\odot}$, $\mu_{\rm gas}=0.08-0.13$, and $t_{\rm dep}=0.65-5.05$ Gyr.
    \item We employ the Kaplan-Meier estimator to compare the redshift independent $\mu_{\rm gas}$ and $t_{\rm dep}$ and upper limits from FRB hosts in the sample and the literature with local galaxies, and find that the gas fraction is statistically different. The significance increases when we remove M81 from the sample, but becomes insignificant again when we only consider the five hosts analyzed here. This suggests that FRBs may have host environments with molecular gas fractions higher than local galaxies, but a larger sample size is necessary to make a more robust comparison.
    \item We briefly compare the kinematic properties of the CO line for HG20180924B with VLT/MUSE observations of the galaxy's H$\alpha$ line and an HST/IR image of the resolved spiral structure of the host.
\end{enumerate}

Future CO surveys of FRB hosts will be able to robustly compare molecular gas properties with local galaxies. When combined with high resolution observations, we will be able to analyze the very local environments of FRBs across the electromagnetic spectrum to help constrain progenitor scenarios.

\section*{Acknowledgements}
The authors would like to thank Nissim Kanekar for many contributions and useful discussions. Authors J.S.C., G.S-M., R.A.J., N.T., J.X.P., T.E., and W.F. as members of the Fast and Fortunate for FRB
Follow-up team, acknowledge support from 
NSF grants AST-1911140, AST-1910471, and AST-2206492, as well as support from NSF REU awards AST-1358980, AST-1757321, and AST-2149985.
This work is supported by the Nantucket Maria Mitchell Association. 

T.E. is supported by NASA through the NASA Hubble Fellowship grant HST-HF2-51504.001-A awarded by the Space Telescope Science Institute, which is operated by the Association of Universities for Research in Astronomy, Inc., for NASA, under contract NAS5-26555.

W.F. gratefully acknowledges support by the David and Lucile Packard Foundation, the Alfred P. Sloan Foundation, and the Research Corporation for Science Advancement through Cottrell Scholar Award \#28284.

R.M.S. acknowledges support through through Australian Research Council Future Fellowship FT190100155 and Discovery Project DP220102305.

This paper makes use of the following ALMA data: ADS/JAO.ALMA $\#$2019.1.01450.S (PI Tejos), $\#$2021.1.01092.S (PI Chittidi).  ALMA is a partnership of ESO (representing its member states), NSF (USA) and NINS (Japan), together with NRC (Canada) and NSC and ASIAA (Taiwan), in cooperation with the Republic of Chile. The Joint ALMA Observatory is operated by ESO, AUI/NRAO and NAOJ. 

Based on observations collected at the European Southern Observatory under ESO programmes 2102.A-5005 (PI: Macquart) and 0104.A-0411 (PI: Tejos).
This research is based on observations made with the NASA/ESA Hubble Space Telescope obtained from the Space Telescope Science Institute, which is operated by the Association of Universities for Research in Astronomy, Inc., under NASA contract NAS 526555. These observations are associated with program No. 15878 with support provided through a grant from the STScI under NASA contract NAS5-26555.

\software{\texttt{astropy} \citep{astropy2013,astropy2018}, \texttt{CASA} \cite[v6.5.3][]{CASA}, \texttt{FRB repository} \citep{frbrepo}, \texttt{lifelines} \citep{Davidson-Pilon2019}, \texttt{matplotlib} \citep{hunter2007}, \texttt{numpy} \citep{harris2020}, \texttt{photutils} \citep{bradley2021}, \texttt{regions} \citep{Regions}, \texttt{scipy} \citep{virtanen2020}, \texttt{spectral-cube} \citep{spectralcube}}

\bibliographystyle{aasjournal}
\bibliography{refs.bib}

\begin{thebibliography}{}
\expandafter\ifx\csname natexlab\endcsname\relax\def\natexlab#1{#1}\fi
\providecommand{\url}[1]{\href{#1}{#1}}
\providecommand{\dodoi}[1]{doi:~\href{http://doi.org/#1}{\nolinkurl{#1}}}
\providecommand{\doeprint}[1]{\href{http://ascl.net/#1}{\nolinkurl{http://ascl.net/#1}}}
\providecommand{\doarXiv}[1]{\href{https://arxiv.org/abs/#1}{\nolinkurl{https://arxiv.org/abs/#1}}}

\bibitem[{{Andreoni} {et~al.}(2020){Andreoni}, {Lu}, {Smith}, {Masci}, {Bellm},
  {Graham}, {Kaplan}, {Kasliwal}, {Kaye}, {Kupfer}, {Laher}, {Mahabal},
  {Nordin}, {Porter}, {Prince}, {Reiley}, {Riddle}, {Van Roestel}, \&
  {Yao}}]{Andreoni2020}
{Andreoni}, I., {Lu}, W., {Smith}, R.~M., {et~al.} 2020, \apjl, 896, L2,
  \dodoi{10.3847/2041-8213/ab94a5}

\bibitem[{Arabsalmani {et~al.}(2019a)Arabsalmani, Roychowdhury, Starkenburg,
  Christensen, Floc'h, Kanekar, Bournaud, Zwaan, Fynbo, M{\o}ller, \&
  Pian}]{Arabsalmani2019a}
Arabsalmani, M., Roychowdhury, S., Starkenburg, T.~K., {et~al.} 2019a, Monthly
  Notices of the Royal Astronomical Society, 485, 5411,
  \dodoi{10.1093/mnras/stz735}

\bibitem[{{Arabsalmani} {et~al.}(2019b){Arabsalmani}, {Roychowdhury}, {Renaud},
  {Cormier}, {Le Floc'h}, {Emsellem}, {Perley}, {Zwaan}, {Bournaud},
  {Arumugam}, \& {M{\o}ller}}]{Arabsalmani2019b}
{Arabsalmani}, M., {Roychowdhury}, S., {Renaud}, F., {et~al.} 2019b, \apj, 882,
  31, \dodoi{10.3847/1538-4357/ab2897}

\bibitem[{{Astropy Collaboration} {et~al.}(2013){Astropy Collaboration},
  {Robitaille}, {Tollerud}, {Greenfield}, {Droettboom}, {Bray}, {Aldcroft},
  {Davis}, {Ginsburg}, {Price-Whelan}, {Kerzendorf}, {Conley}, {Crighton},
  {Barbary}, {Muna}, {Ferguson}, {Grollier}, {Parikh}, {Nair}, {Unther},
  {Deil}, {Woillez}, {Conseil}, {Kramer}, {Turner}, {Singer}, {Fox}, {Weaver},
  {Zabalza}, {Edwards}, {Azalee Bostroem}, {Burke}, {Casey}, {Crawford},
  {Dencheva}, {Ely}, {Jenness}, {Labrie}, {Lim}, {Pierfederici}, {Pontzen},
  {Ptak}, {Refsdal}, {Servillat}, \& {Streicher}}]{astropy2013}
{Astropy Collaboration}, {Robitaille}, T.~P., {Tollerud}, E.~J., {et~al.} 2013,
  \aap, 558, A33, \dodoi{10.1051/0004-6361/201322068}

\bibitem[{{Astropy Collaboration} {et~al.}(2018){Astropy Collaboration},
  {Price-Whelan}, {Sip{H{o}}cz}, {G{"u}nther}, {Lim}, {Crawford}, {Conseil},
  {Shupe}, {Craig}, {Dencheva}, {Ginsburg}, {Vand erPlas}, {Bradley},
  {P{'e}rez-Su{'a}rez}, {de Val-Borro}, {Aldcroft}, {Cruz}, {Robitaille},
  {Tollerud}, {Ardelean}, {Babej}, {Bach}, {Bachetti}, {Bakanov}, {Bamford},
  {Barentsen}, {Barmby}, {Baumbach}, {Berry}, {Biscani}, {Boquien}, {Bostroem},
  {Bouma}, {Brammer}, {Bray}, {Breytenbach}, {Buddelmeijer}, {Burke},
  {Calderone}, {Cano Rodr{'i}guez}, {Cara}, {Cardoso}, {Cheedella}, {Copin},
  {Corrales}, {Crichton}, {D'Avella}, {Deil}, {Depagne}, {Dietrich}, {Donath},
  {Droettboom}, {Earl}, {Erben}, {Fabbro}, {Ferreira}, {Finethy}, {Fox},
  {Garrison}, {Gibbons}, {Goldstein}, {Gommers}, {Greco}, {Greenfield},
  {Groener}, {Grollier}, {Hagen}, {Hirst}, {Homeier}, {Horton}, {Hosseinzadeh},
  {Hu}, {Hunkeler}, {Ivezi{'c}}, {Jain}, {Jenness}, {Kanarek}, {Kendrew},
  {Kern}, {Kerzendorf}, {Khvalko}, {King}, {Kirkby}, {Kulkarni}, {Kumar},
  {Lee}, {Lenz}, {Littlefair}, {Ma}, {Macleod}, {Mastropietro}, {McCully},
  {Montagnac}, {Morris}, {Mueller}, {Mumford}, {Muna}, {Murphy}, {Nelson},
  {Nguyen}, {Ninan}, {N{"o}the}, {Ogaz}, {Oh}, {Parejko}, {Parley}, {Pascual},
  {Patil}, {Patil}, {Plunkett}, {Prochaska}, {Rastogi}, {Reddy Janga},
  {Sabater}, {Sakurikar}, {Seifert}, {Sherbert}, {Sherwood-Taylor}, {Shih},
  {Sick}, {Silbiger}, {Singanamalla}, {Singer}, {Sladen}, {Sooley},
  {Sornarajah}, {Streicher}, {Teuben}, {Thomas}, {Tremblay}, {Turner},
  {Terr{'o}n}, {van Kerkwijk}, {de la Vega}, {Watkins}, {Weaver}, {Whitmore},
  {Woillez}, {Zabalza}, \& {Astropy Contributors}}]{astropy2018}
{Astropy Collaboration}, {Price-Whelan}, A.~M., {Sip{H{o}}cz}, B.~M., {et~al.}
  2018, aj, 156, 123, \dodoi{10.3847/1538-3881/aabc4f}

\bibitem[{{Bacon} {et~al.}(2010){Bacon}, {Accardo}, {Adjali}, {Anwand},
  {Bauer}, {Biswas}, {Blaizot}, {Boudon}, {Brau-Nogue}, {Brinchmann},
  {Caillier}, {Capoani}, {Carollo}, {Contini}, {Couderc}, {Daguis{\'e}},
  {Deiries}, {Delabre}, {Dreizler}, {Dubois}, {Dupieux}, {Dupuy}, {Emsellem},
  {Fechner}, {Fleischmann}, {Fran{\c{c}}ois}, {Gallou}, {Gharsa}, {Glindemann},
  {Gojak}, {Guiderdoni}, {Hansali}, {Hahn}, {Jarno}, {Kelz}, {Koehler},
  {Kosmalski}, {Laurent}, {Le Floch}, {Lilly}, {Lizon}, {Loupias}, {Manescau},
  {Monstein}, {Nicklas}, {Olaya}, {Pares}, {Pasquini}, {P{\'e}contal-Rousset},
  {Pell{\'o}}, {Petit}, {Popow}, {Reiss}, {Remillieux}, {Renault}, {Roth},
  {Rupprecht}, {Serre}, {Schaye}, {Soucail}, {Steinmetz}, {Streicher}, {Stuik},
  {Valentin}, {Vernet}, {Weilbacher}, {Wisotzki}, \& {Yerle}}]{MUSE}
{Bacon}, R., {Accardo}, M., {Adjali}, L., {et~al.} 2010, in Society of
  Photo-Optical Instrumentation Engineers (SPIE) Conference Series, Vol. 7735,
  Ground-based and Airborne Instrumentation for Astronomy III, ed. I.~S.
  {McLean}, S.~K. {Ramsay}, \& H.~{Takami}, 773508, \dodoi{10.1117/12.856027}

\bibitem[{{Bandura} {et~al.}(2014){Bandura}, {Addison}, {Amiri}, {Bond},
  {Campbell-Wilson}, {Connor}, {Cliche}, {Davis}, {Deng}, {Denman}, {Dobbs},
  {Fandino}, {Gibbs}, {Gilbert}, {Halpern}, {Hanna}, {Hincks}, {Hinshaw},
  {H{\"o}fer}, {Klages}, {Landecker}, {Masui}, {Mena Parra}, {Newburgh}, {Pen},
  {Peterson}, {Recnik}, {Shaw}, {Sigurdson}, {Sitwell}, {Smecher}, {Smegal},
  {Vanderlinde}, \& {Wiebe}}]{chime2014}
{Bandura}, K., {Addison}, G.~E., {Amiri}, M., {et~al.} 2014, in procspie, Vol.
  9145, Ground-based and Airborne Telescopes V, 914522.
\newblock \doarXiv{1406.2288}

\bibitem[{Bannister {et~al.}(2019)Bannister, Deller, Phillips, Macquart,
  Prochaska, Tejos, Ryder, Sadler, Shannon, Simha, \& et~al.}]{Bannister2019}
Bannister, K.~W., Deller, A.~T., Phillips, C., {et~al.} 2019, Science, 365,
  565–570, \dodoi{10.1126/science.aaw5903}

\bibitem[{{Bhandari} {et~al.}(2020{\natexlab{a}}){Bhandari}, {Sadler},
  {Prochaska}, {Simha}, {Ryder}, {Marnoch}, {Bannister}, {Macquart}, {Flynn},
  {Shannon}, {Tejos}, {Corro-Guerra}, {Day}, {Deller}, {Ekers}, {Lopez},
  {Mahony}, {Nu{\~n}ez}, \& {Phillips}}]{Bhandari2020a}
{Bhandari}, S., {Sadler}, E.~M., {Prochaska}, J.~X., {et~al.}
  2020{\natexlab{a}}, \apjl, 895, L37, \dodoi{10.3847/2041-8213/ab672e}

\bibitem[{{Bhandari} {et~al.}(2020{\natexlab{b}}){Bhandari}, {Bannister},
  {Lenc}, {Cho}, {Ekers}, {Day}, {Deller}, {Flynn}, {James}, {Macquart},
  {Mahony}, {Marnoch}, {Moss}, {Phillips}, {Prochaska}, {Qiu}, {Ryder},
  {Shannon}, {Tejos}, \& {Wong}}]{Bhandari2020b}
{Bhandari}, S., {Bannister}, K.~W., {Lenc}, E., {et~al.} 2020{\natexlab{b}},
  \apjl, 901, L20, \dodoi{10.3847/2041-8213/abb462}

\bibitem[{{Bhardwaj} {et~al.}(2021){Bhardwaj}, {Gaensler}, {Kaspi},
  {Landecker}, {Mckinven}, {Michilli}, {Pleunis}, {Tendulkar}, {Andersen},
  {Boyle}, {Cassanelli}, {Chawla}, {Cook}, {Dobbs}, {Fonseca}, {Kaczmarek},
  {Leung}, {Masui}, {Mnchmeyer}, {Ng}, {Rafiei-Ravandi}, {Scholz}, {Shin},
  {Smith}, {Stairs}, \& {Zwaniga}}]{Bhardwaj2021}
{Bhardwaj}, M., {Gaensler}, B.~M., {Kaspi}, V.~M., {et~al.} 2021, \apjl, 910,
  L18, \dodoi{10.3847/2041-8213/abeaa6}

\bibitem[{{Bolatto} {et~al.}(2013){Bolatto}, {Wolfire}, \&
  {Leroy}}]{Bolatto2013}
{Bolatto}, A.~D., {Wolfire}, M., \& {Leroy}, A.~K. 2013, \araa, 51, 207,
  \dodoi{10.1146/annurev-astro-082812-140944}

\bibitem[{{Bower} {et~al.}(2018){Bower}, {Rao}, {Krips}, {Maddox}, {Bassa},
  {Adams}, {Law}, {Tendulkar}, {van Langevelde}, {Paragi}, {Butler}, \&
  {Chatterjee}}]{Bower2018}
{Bower}, G.~C., {Rao}, R., {Krips}, M., {et~al.} 2018, \aj, 155, 227,
  \dodoi{10.3847/1538-3881/aabc5a}

\bibitem[{{Bradley} {et~al.}(2021){Bradley}, {Sip{\H{o}}cz}, {Robitaille},
  {Tollerud}, {Vin{\'\i}cius}, {Deil}, {Barbary}, {Wilson}, {Busko}, {Donath},
  {G{\"u}nther}, {Cara}, {Conseil}, {Bostroem}, {Droettboom}, {Bray},
  {Krachyon}, {Lim}, {Andersen Bratholm}, {Barentsen}, {Craig}, {Rathi},
  {Pascual}, {Perren}, {Georgiev}, {De Val-Borro}, {Kerzendorf}, {Bach},
  {Quint}, \& {Souchereau}}]{bradley2021}
{Bradley}, L., {Sip{\H{o}}cz}, B., {Robitaille}, T., {et~al.} 2021,
  {astropy/photutils: 1.1.0}, 1.1.0, Zenodo,  Zenodo,
  \dodoi{10.5281/zenodo.4624996}

\bibitem[{Bradley {et~al.}(2022)Bradley, Deil, Ginsburg, Patra, Robitaille,
  Sipőcz, King, Lim, Homeier, Singer, de~Val-Borro, Jenness, Baumann,
  Gondhalekar, Donath, Tollerud, Lee, Leinweber, \& Vinícius}]{Regions}
Bradley, L., Deil, C., Ginsburg, A., {et~al.} 2022, astropy/regions: v0.7,
  v0.7,  Zenodo, \dodoi{10.5281/zenodo.7259631}

\bibitem[{{CASA Team} {et~al.}(2022){CASA Team}, {Bean}, {Bhatnagar}, {Castro},
  {Donovan Meyer}, {Emonts}, {Garcia}, {Garwood}, {Golap}, {Gonzalez Villalba},
  {Harris}, {Hayashi}, {Hoskins}, {Hsieh}, {Jagannathan}, {Kawasaki},
  {Keimpema}, {Kettenis}, {Lopez}, {Marvil}, {Masters}, {McNichols},
  {Mehringer}, {Miel}, {Moellenbrock}, {Montesino}, {Nakazato}, {Ott}, {Petry},
  {Pokorny}, {Raba}, {Rau}, {Schiebel}, {Schweighart}, {Sekhar}, {Shimada},
  {Small}, {Steeb}, {Sugimoto}, {Suoranta}, {Tsutsumi}, {van Bemmel},
  {Verkouter}, {Wells}, {Xiong}, {Szomoru}, {Griffith}, {Glendenning}, \&
  {Kern}}]{CASA}
{CASA Team}, {Bean}, B., {Bhatnagar}, S., {et~al.} 2022, \pasp, 134, 114501,
  \dodoi{10.1088/1538-3873/ac9642}

\bibitem[{{Chen} {et~al.}(2020){Chen}, {Ravi}, \& {Lu}}]{Chen2020}
{Chen}, G., {Ravi}, V., \& {Lu}, W. 2020, \apj, 897, 146,
  \dodoi{10.3847/1538-4357/ab982b}

\bibitem[{{Chen} {et~al.}(2023){Chen}, {Ivison}, {Zwaan}, {Smail}, {Klitsch},
  {P{\'e}roux}, {Popping}, {Biggs}, {Szakacs}, {Hamanowicz}, \&
  {Lagos}}]{Chen2023}
{Chen}, J., {Ivison}, R.~J., {Zwaan}, M.~A., {et~al.} 2023, \mnras, 518, 1378,
  \dodoi{10.1093/mnras/stac2989}

\bibitem[{{CHIME/FRB Collaboration} {et~al.}(2019){CHIME/FRB Collaboration},
  {Andersen}, {Bandura}, {Bhardwaj}, {Boubel}, {Boyce}, {Boyle}, {Brar},
  {Cassanelli}, {Chawla}, {Cubranic}, {Deng}, {Dobbs}, {Fandino}, {Fonseca},
  {Gaensler}, {Gilbert}, {Giri}, {Good}, {Halpern}, {Hill}, {Hinshaw},
  {H{\"o}fer}, {Josephy}, {Kaspi}, {Kothes}, {Landecker}, {Lang}, {Li}, {Lin},
  {Masui}, {Mena-Parra}, {Merryfield}, {Mckinven}, {Michilli}, {Milutinovic},
  {Naidu}, {Newburgh}, {Ng}, {Patel}, {Pen}, {Pinsonneault-Marotte}, {Pleunis},
  {Rafiei-Ravandi}, {Rahman}, {Ransom}, {Renard}, {Scholz}, {Siegel}, {Singh},
  {Smith}, {Stairs}, {Tendulkar}, {Tretyakov}, {Vanderlinde}, {Yadav}, \&
  {Zwaniga}}]{chime2019}
{CHIME/FRB Collaboration}, {Andersen}, B.~C., {Bandura}, K., {et~al.} 2019,
  \apjl, 885, L24, \dodoi{10.3847/2041-8213/ab4a80}

\bibitem[{{CHIME/FRB Collaboration} {et~al.}(2020){CHIME/FRB Collaboration},
  {Andersen}, {Bandura}, {Bhardwaj}, {Bij}, {Boyce}, {Boyle}, {Brar},
  {Cassanelli}, {Chawla}, {Chen}, {Cliche}, {Cook}, {Cubranic}, {Curtin},
  {Denman}, {Dobbs}, {Dong}, {Fandino}, {Fonseca}, {Gaensler}, {Giri}, {Good},
  {Halpern}, {Hill}, {Hinshaw}, {H{\"o}fer}, {Josephy}, {Kania}, {Kaspi},
  {Landecker}, {Leung}, {Li}, {Lin}, {Masui}, {McKinven}, {Mena-Parra},
  {Merryfield}, {Meyers}, {Michilli}, {Milutinovic}, {Mirhosseini},
  {M{\"u}nchmeyer}, {Naidu}, {Newburgh}, {Ng}, {Patel}, {Pen},
  {Pinsonneault-Marotte}, {Pleunis}, {Quine}, {Rafiei-Ravandi}, {Rahman},
  {Ransom}, {Renard}, {Sanghavi}, {Scholz}, {Shaw}, {Shin}, {Siegel}, {Singh},
  {Smegal}, {Smith}, {Stairs}, {Tan}, {Tendulkar}, {Tretyakov}, {Vanderlinde},
  {Wang}, {Wulf}, \& {Zwaniga}}]{CHIME2020magnetar}
{CHIME/FRB Collaboration}, {Andersen}, B.~C., {Bandura}, K.~M., {et~al.} 2020,
  \nat, 587, 54, \dodoi{10.1038/s41586-020-2863-y}

\bibitem[{{CHIME/FRB Collaboration} {et~al.}(2021){CHIME/FRB Collaboration},
  {Amiri}, {Andersen}, {Bandura}, {Berger}, {Bhardwaj}, {Boyce}, {Boyle},
  {Brar}, {Breitman}, {Cassanelli}, {Chawla}, {Chen}, {Cliche}, {Cook},
  {Cubranic}, {Curtin}, {Deng}, {Dobbs}, {Dong}, {Eadie}, {Fandino}, {Fonseca},
  {Gaensler}, {Giri}, {Good}, {Halpern}, {Hill}, {Hinshaw}, {Josephy},
  {Kaczmarek}, {Kader}, {Kania}, {Kaspi}, {Landecker}, {Lang}, {Leung}, {Li},
  {Lin}, {Masui}, {McKinven}, {Mena-Parra}, {Merryfield}, {Meyers}, {Michilli},
  {Milutinovic}, {Mirhosseini}, {M{\"u}nchmeyer}, {Naidu}, {Newburgh}, {Ng},
  {Patel}, {Pen}, {Petroff}, {Pinsonneault-Marotte}, {Pleunis},
  {Rafiei-Ravandi}, {Rahman}, {Ransom}, {Renard}, {Sanghavi}, {Scholz}, {Shaw},
  {Shin}, {Siegel}, {Sikora}, {Singh}, {Smith}, {Stairs}, {Tan}, {Tendulkar},
  {Vanderlinde}, {Wang}, {Wulf}, \& {Zwaniga}}]{CHIME2021}
{CHIME/FRB Collaboration}, {Amiri}, M., {Andersen}, B.~C., {et~al.} 2021,
  \apjs, 257, 59, \dodoi{10.3847/1538-4365/ac33ab}

\bibitem[{Chittidi {et~al.}(2021)Chittidi, Simha, Mannings, Prochaska, Ryder,
  Rafelski, Neeleman, Macquart, Tejos, Jorgenson, \& et~al.}]{Chittidi2021}
Chittidi, J.~S., Simha, S., Mannings, A., {et~al.} 2021, The Astrophysical
  Journal, 922, 173, \dodoi{10.3847/1538-4357/ac2818}

\bibitem[{Cordes \& Chatterjee(2019)}]{CordesChatterjee2019}
Cordes, J.~M., \& Chatterjee, S. 2019, Annual Review of Astronomy and
  Astrophysics, 57, 417, \dodoi{10.1146/annurev-astro-091918-104501}

\bibitem[{Davidson-Pilon(2019)}]{Davidson-Pilon2019}
Davidson-Pilon, C. 2019, Journal of Open Source Software, 4, 1317,
  \dodoi{10.21105/joss.01317}

\bibitem[{Day {et~al.}(2021)Day, Deller, James, Lenc, Bhandari, Shannon, \&
  Bannister}]{Day2021}
Day, C.~K., Deller, A.~T., James, C.~W., {et~al.} 2021, Publications of the
  Astronomical Society of Australia, 38, \dodoi{10.1017/pasa.2021.40}

\bibitem[{{Eftekhari} \& {Berger}(2017)}]{Eftekhari2017}
{Eftekhari}, T., \& {Berger}, E. 2017, \apj, 849, 162,
  \dodoi{10.3847/1538-4357/aa90b9}

\bibitem[{{Galbany} {et~al.}(2017){Galbany}, {Mora}, {Gonz{\'a}lez-Gait{\'a}n},
  {Bolatto}, {Dannerbauer}, {L{\'o}pez-S{\'a}nchez}, {Maeda}, {P{\'e}rez},
  {P{\'e}rez-Torres}, {S{\'a}nchez}, {Wong}, {Badenes}, {Blitz}, {Marino},
  {Utomo}, \& {Van de Ven}}]{Galbany2017}
{Galbany}, L., {Mora}, L., {Gonz{\'a}lez-Gait{\'a}n}, S., {et~al.} 2017,
  \mnras, 468, 628, \dodoi{10.1093/mnras/stx367}

\bibitem[{{Genzel} {et~al.}(2015){Genzel}, {Tacconi}, {Lutz}, {Saintonge},
  {Berta}, {Magnelli}, {Combes}, {Garc{\'\i}a-Burillo}, {Neri}, {Bolatto},
  {Contini}, {Lilly}, {Boissier}, {Boone}, {Bouch{\'e}}, {Bournaud}, {Burkert},
  {Carollo}, {Colina}, {Cooper}, {Cox}, {Feruglio}, {F{\"o}rster Schreiber},
  {Freundlich}, {Gracia-Carpio}, {Juneau}, {Kovac}, {Lippa}, {Naab}, {Salome},
  {Renzini}, {Sternberg}, {Walter}, {Weiner}, {Weiss}, \& {Wuyts}}]{genzel2015}
{Genzel}, R., {Tacconi}, L.~J., {Lutz}, D., {et~al.} 2015, \apj, 800, 20,
  \dodoi{10.1088/0004-637X/800/1/20}

\bibitem[{Gordon {et~al.}(2023)Gordon, fai Fong, Kilpatrick, Eftekhari, Leja,
  Prochaska, Nugent, Bhandari, Blanchard, Caleb, Day, Deller, Dong, Glowacki,
  Gourdji, Mannings, Mahoney, Marnoch, Miller, Paterson, Rastinejad, Ryder,
  Sadler, Scott, Sears, Shannon, Simha, Stappers, \& Tejos}]{Gordon2023}
Gordon, A.~C., fai Fong, W., Kilpatrick, C.~D., {et~al.} 2023.
\newblock \doarXiv{2302.05465}

\bibitem[{{Hagstotz} {et~al.}(2022){Hagstotz}, {Reischke}, \&
  {Lilow}}]{Hagstotz2022}
{Hagstotz}, S., {Reischke}, R., \& {Lilow}, R. 2022, \mnras, 511, 662,
  \dodoi{10.1093/mnras/stac077}

\bibitem[{{Hardy} {et~al.}(2017){Hardy}, {Dhillon}, {Spitler}, {Littlefair},
  {Ashley}, {De Cia}, {Green}, {Jaroenjittichai}, {Keane}, {Kerry}, {Kramer},
  {Malesani}, {Marsh}, {Parsons}, {Possenti}, {Rattanasoon}, \&
  {Sahman}}]{Hardy2017}
{Hardy}, L.~K., {Dhillon}, V.~S., {Spitler}, L.~G., {et~al.} 2017, \mnras, 472,
  2800, \dodoi{10.1093/mnras/stx2153}

\bibitem[{{Harris} {et~al.}(2020){Harris}, {Millman}, {van der Walt},
  {Gommers}, {Virtanen}, {Cournapeau}, {Wieser}, {Taylor}, {Berg}, {Smith},
  {Kern}, {Picus}, {Hoyer}, {van Kerkwijk}, {Brett}, {Haldane}, {del R{\'\i}o},
  {Wiebe}, {Peterson}, {G{\'e}rard-Marchant}, {Sheppard}, {Reddy}, {Weckesser},
  {Abbasi}, {Gohlke}, \& {Oliphant}}]{harris2020}
{Harris}, C.~R., {Millman}, K.~J., {van der Walt}, S.~J., {et~al.} 2020, \nat,
  585, 357, \dodoi{10.1038/s41586-020-2649-2}

\bibitem[{{Hatsukade} {et~al.}(2022){Hatsukade}, {Hashimoto}, {Niino}, \&
  {Hsu}}]{Hatsukade2022}
{Hatsukade}, B., {Hashimoto}, T., {Niino}, Y., \& {Hsu}, T.-Y. 2022, \apjl,
  940, L34, \dodoi{10.3847/2041-8213/ac9f39}

\bibitem[{{Hatsukade} {et~al.}(2020{\natexlab{a}}){Hatsukade}, {Ohta},
  {Hashimoto}, {Kohno}, {Nakanishi}, {Niino}, \& {Tamura}}]{Hatsukade2020_grb}
{Hatsukade}, B., {Ohta}, K., {Hashimoto}, T., {et~al.} 2020{\natexlab{a}},
  \apj, 892, 42, \dodoi{10.3847/1538-4357/ab7992}

\bibitem[{{Hatsukade} {et~al.}(2020{\natexlab{b}}){Hatsukade},
  {Morokuma-Matsui}, {Hayashi}, {Tominaga}, {Tamura}, {Niinuma}, {Motogi},
  {Morokuma}, \& {Matsuda}}]{Hatsukade2020_slsn}
{Hatsukade}, B., {Morokuma-Matsui}, K., {Hayashi}, M., {et~al.}
  2020{\natexlab{b}}, \pasj, 72, L6, \dodoi{10.1093/pasj/psaa052}

\bibitem[{Heintz {et~al.}(2020)Heintz, Prochaska, Simha, Platts, Fong, Tejos,
  Ryder, Aggerwal, Bhandari, Day, \& et~al.}]{Heintz2020}
Heintz, K.~E., Prochaska, J.~X., Simha, S., {et~al.} 2020, The Astrophysical
  Journal, 903, 152, \dodoi{10.3847/1538-4357/abb6fb}

\bibitem[{{Hiramatsu} {et~al.}(2022){Hiramatsu}, {Berger}, {Metzger}, {Gomez},
  {Bieryla}, {Arcavi}, {Howell}, {Mckinven}, \& {Tominaga}}]{Hiramatsu2022}
{Hiramatsu}, D., {Berger}, E., {Metzger}, B.~D., {et~al.} 2022, arXiv e-prints,
  arXiv:2211.03974, \dodoi{10.48550/arXiv.2211.03974}

\bibitem[{{Hsu} {et~al.}(2023){Hsu}, {Hashimoto}, {Hatsukade}, {Goto}, {Wang},
  {Ling}, {Ho}, \& {Uno}}]{Hsu2023}
{Hsu}, T.-Y., {Hashimoto}, T., {Hatsukade}, B., {et~al.} 2023, \mnras, 519,
  2030, \dodoi{10.1093/mnras/stac3655}

\bibitem[{{Hunter}(2007)}]{hunter2007}
{Hunter}, J.~D. 2007, Computing in Science and Engineering, 9, 90,
  \dodoi{10.1109/MCSE.2007.55}

\bibitem[{{James} {et~al.}(2022){James}, {Ghosh}, {Prochaska}, {Bannister},
  {Bhandari}, {Day}, {Deller}, {Glowacki}, {Gordon}, {Heintz}, {Marnoch},
  {Ryder}, {Scott}, {Shannon}, \& {Tejos}}]{James2022}
{James}, C.~W., {Ghosh}, E.~M., {Prochaska}, J.~X., {et~al.} 2022, \mnras, 516,
  4862, \dodoi{10.1093/mnras/stac2524}

\bibitem[{Kaplan \& Meier(1958)}]{KaplanMeier1958}
Kaplan, E.~L., \& Meier, P. 1958, Journal of the American Statistical
  Association, 53, 457, \dodoi{10.1080/01621459.1958.10501452}

\bibitem[{{Kaur} {et~al.}(2022){Kaur}, {Kanekar}, \& {Prochaska}}]{Kaur2022}
{Kaur}, B., {Kanekar}, N., \& {Prochaska}, J.~X. 2022, \apjl, 925, L20,
  \dodoi{10.3847/2041-8213/ac4ca8}

\bibitem[{{Kilpatrick} {et~al.}(2021){Kilpatrick}, {Burchett}, {Jones},
  {Margalit}, {McMillan}, {Fong}, {Heintz}, {Tejos}, \&
  {Escorial}}]{Kilpatrick2021}
{Kilpatrick}, C.~D., {Burchett}, J.~N., {Jones}, D.~O., {et~al.} 2021, \apjl,
  907, L3, \dodoi{10.3847/2041-8213/abd560}

\bibitem[{{Kirsten} {et~al.}(2022){Kirsten}, {Marcote}, {Nimmo}, {Hessels},
  {Bhardwaj}, {Tendulkar}, {Keimpema}, {Yang}, {Snelders}, {Scholz},
  {Pearlman}, {Law}, {Peters}, {Giroletti}, {Paragi}, {Bassa}, {Hewitt},
  {Bach}, {Bezrukovs}, {Burgay}, {Buttaccio}, {Conway}, {Corongiu}, {Feiler},
  {Forss{\'e}n}, {Gawro{\'n}ski}, {Karuppusamy}, {Kharinov}, {Lindqvist},
  {Maccaferri}, {Melnikov}, {Ould-Boukattine}, {Possenti}, {Surcis}, {Wang},
  {Yuan}, {Aggarwal}, {Anna-Thomas}, {Bower}, {Blaauw}, {Burke-Spolaor},
  {Cassanelli}, {Clarke}, {Fonseca}, {Gaensler}, {Gopinath}, {Kaspi}, {Kassim},
  {Lazio}, {Leung}, {Li}, {Lin}, {Masui}, {Mckinven}, {Michilli}, {Mikhailov},
  {Ng}, {Orbidans}, {Pen}, {Petroff}, {Rahman}, {Ransom}, {Shin}, {Smith},
  {Stairs}, \& {Vlemmings}}]{Kirsten2022}
{Kirsten}, F., {Marcote}, B., {Nimmo}, K., {et~al.} 2022, \nat, 602, 585,
  \dodoi{10.1038/s41586-021-04354-w}

\bibitem[{Koch {et~al.}(2019)Koch, Ginsburg, Robitaille, Beaumont, Rosolowsky,
  \& Leroy}]{spectralcube}
Koch, E., Ginsburg, A., Robitaille, T., {et~al.} 2019,
  radio-astro-tools/spectral-cube: Release v0.4.5, v0.4.5,  Zenodo,
  \dodoi{10.5281/zenodo.3558614}

\bibitem[{{Kocz} {et~al.}(2019){Kocz}, {Ravi}, {Catha}, {D'Addario},
  {Hallinan}, {Hobbs}, {Kulkarni}, {Shi}, {Vedantham}, {Weinreb}, \&
  {Woody}}]{Kocz2019}
{Kocz}, J., {Ravi}, V., {Catha}, M., {et~al.} 2019, \mnras, 489, 919,
  \dodoi{10.1093/mnras/stz2219}

\bibitem[{{Lamperti} {et~al.}(2020){Lamperti}, {Saintonge}, {Koss}, {Viti},
  {Wilson}, {He}, {Shimizu}, {Greve}, {Mushotzky}, {Treister}, {Kramer},
  {Sanders}, {Schawinski}, \& {Tacconi}}]{Lamperti2020}
{Lamperti}, I., {Saintonge}, A., {Koss}, M., {et~al.} 2020, \apj, 889, 103,
  \dodoi{10.3847/1538-4357/ab6221}

\bibitem[{{Law} {et~al.}(2018){Law}, {Bower}, {Burke-Spolaor}, {Butler},
  {Demorest}, {Halle}, {Khudikyan}, {Lazio}, {Pokorny}, {Robnett}, \&
  {Rupen}}]{Law2018}
{Law}, C.~J., {Bower}, G.~C., {Burke-Spolaor}, S., {et~al.} 2018, \apjs, 236,
  8, \dodoi{10.3847/1538-4365/aab77b}

\bibitem[{{Lorimer} {et~al.}(2007){Lorimer}, {Bailes}, {McLaughlin},
  {Narkevic}, \& {Crawford}}]{lorimer2007}
{Lorimer}, D.~R., {Bailes}, M., {McLaughlin}, M.~A., {Narkevic}, D.~J., \&
  {Crawford}, F. 2007, Science, 318, 777, \dodoi{10.1126/science.1147532}

\bibitem[{Macquart {et~al.}(2010)Macquart, Bailes, Bhat, Bower, Bunton,
  Chatterjee, Colegate, Cordes, D'Addario, Deller, \& et~al.}]{Macquart2010}
Macquart, J.-P., Bailes, M., Bhat, N. D.~R., {et~al.} 2010, Publications of the
  Astronomical Society of Australia, 27, 272–282, \dodoi{10.1071/AS09082}

\bibitem[{Macquart {et~al.}(2020)Macquart, Prochaska, McQuinn, Bannister,
  Bhandari, Day, Deller, Ekers, James, Marnoch, \& et~al.}]{Macquart2020}
Macquart, J.-P., Prochaska, J.~X., McQuinn, M., {et~al.} 2020, Nature, 581,
  391–395, \dodoi{10.1038/s41586-020-2300-2}

\bibitem[{Mannings {et~al.}(2020)Mannings, fai Fong, Simha, Prochaska,
  Rafelski, Kilpatrick, Tejos, Heintz, Bhandari, Day, Deller, Ryder, Shannon,
  \& Tendulkar}]{Mannings2020}
Mannings, A.~G., fai Fong, W., Simha, S., {et~al.} 2020, A High-Resolution View
  of Fast Radio Burst Host Environments.
\newblock \doarXiv{2012.11617}

\bibitem[{{Marcote} {et~al.}(2020){Marcote}, {Nimmo}, {Hessels}, {Tendulkar},
  {Bassa}, {Paragi}, {Keimpema}, {Bhardwaj}, {Karuppusamy}, {Kaspi}, {Law},
  {Michilli}, {Aggarwal}, {Andersen}, {Archibald}, {Bandura}, {Bower}, {Boyle},
  {Brar}, {Burke-Spolaor}, {Butler}, {Cassanelli}, {Chawla}, {Demorest},
  {Dobbs}, {Fonseca}, {Giri}, {Good}, {Gourdji}, {Josephy}, {Kirichenko},
  {Kirsten}, {Landecker}, {Lang}, {Lazio}, {Li}, {Lin}, {Linford}, {Masui},
  {Mena-Parra}, {Naidu}, {Ng}, {Patel}, {Pen}, {Pleunis}, {Rafiei-Ravandi},
  {Rahman}, {Renard}, {Scholz}, {Siegel}, {Smith}, {Stairs}, {Vanderlinde}, \&
  {Zwaniga}}]{Marcote2020}
{Marcote}, B., {Nimmo}, K., {Hessels}, J.~W.~T., {et~al.} 2020, \nat, 577, 190,
  \dodoi{10.1038/s41586-019-1866-z}

\bibitem[{{Marnoch} {et~al.}(2020){Marnoch}, {Ryder}, {Bannister}, {Bhandari},
  {Day}, {Deller}, {Macquart}, {McDermid}, {Xavier Prochaska}, {Qiu}, {Sadler},
  {Shannon}, \& {Tejos}}]{Marnoch2020}
{Marnoch}, L., {Ryder}, S.~D., {Bannister}, K.~W., {et~al.} 2020, \aap, 639,
  A119, \dodoi{10.1051/0004-6361/202038076}

\bibitem[{McConnell {et~al.}(2016)McConnell, Allison, Bannister, Bell, Bignall,
  Chippendale, Edwards, Harvey-Smith, Hegarty, Heywood, \&
  et~al.}]{Mcconnelletal2016}
McConnell, D., Allison, J.~R., Bannister, K., {et~al.} 2016, Publications of
  the Astronomical Society of Australia, 33, e042, \dodoi{10.1017/pasa.2016.37}

\bibitem[{{N{\'u}{\~n}ez} {et~al.}(2021){N{\'u}{\~n}ez}, {Tejos}, {Pignata},
  {Kilpatrick}, {Prochaska}, {Heintz}, {Bannister}, {Bhandari}, {Day},
  {Deller}, {Flynn}, {Mahony}, {Majewski}, {Marnoch}, {Qiu}, {Ryder}, \&
  {Shannon}}]{Nunez2021}
{N{\'u}{\~n}ez}, C., {Tejos}, N., {Pignata}, G., {et~al.} 2021, \aap, 653,
  A119, \dodoi{10.1051/0004-6361/202141110}

\bibitem[{{Petroff} {et~al.}(2022){Petroff}, {Hessels}, \&
  {Lorimer}}]{petroff2022}
{Petroff}, E., {Hessels}, J.~W.~T., \& {Lorimer}, D.~R. 2022, \aapr, 30, 2,
  \dodoi{10.1007/s00159-022-00139-w}

\bibitem[{{Planck Collaboration} {et~al.}(2016){Planck Collaboration}, {Ade},
  {Aghanim}, {Arnaud}, {Ashdown}, {Aumont}, {Baccigalupi}, {Banday},
  {Barreiro}, {Bartlett}, {Bartolo}, {Battaner}, {Battye}, {Benabed},
  {Beno{\^\i}t}, {Benoit-L{\'e}vy}, {Bernard}, {Bersanelli}, {Bielewicz},
  {Bock}, {Bonaldi}, {Bonavera}, {Bond}, {Borrill}, {Bouchet}, {Boulanger},
  {Bucher}, {Burigana}, {Butler}, {Calabrese}, {Cardoso}, {Catalano},
  {Challinor}, {Chamballu}, {Chary}, {Chiang}, {Chluba}, {Christensen},
  {Church}, {Clements}, {Colombi}, {Colombo}, {Combet}, {Coulais}, {Crill},
  {Curto}, {Cuttaia}, {Danese}, {Davies}, {Davis}, {de Bernardis}, {de Rosa},
  {de Zotti}, {Delabrouille}, {D{\'e}sert}, {Di Valentino}, {Dickinson},
  {Diego}, {Dolag}, {Dole}, {Donzelli}, {Dor{\'e}}, {Douspis}, {Ducout},
  {Dunkley}, {Dupac}, {Efstathiou}, {Elsner}, {En{\ss}lin}, {Eriksen},
  {Farhang}, {Fergusson}, {Finelli}, {Forni}, {Frailis}, {Fraisse},
  {Franceschi}, {Frejsel}, {Galeotta}, {Galli}, {Ganga}, {Gauthier}, {Gerbino},
  {Ghosh}, {Giard}, {Giraud-H{\'e}raud}, {Giusarma}, {Gjerl{\o}w},
  {Gonz{\'a}lez-Nuevo}, {G{\'o}rski}, {Gratton}, {Gregorio}, {Gruppuso},
  {Gudmundsson}, {Hamann}, {Hansen}, {Hanson}, {Harrison}, {Helou},
  {Henrot-Versill{\'e}}, {Hern{\'a}ndez-Monteagudo}, {Herranz}, {Hildebrandt},
  {Hivon}, {Hobson}, {Holmes}, {Hornstrup}, {Hovest}, {Huang}, {Huffenberger},
  {Hurier}, {Jaffe}, {Jaffe}, {Jones}, {Juvela}, {Keih{\"a}nen}, {Keskitalo},
  {Kisner}, {Kneissl}, {Knoche}, {Knox}, {Kunz}, {Kurki-Suonio}, {Lagache},
  {L{\"a}hteenm{\"a}ki}, {Lamarre}, {Lasenby}, {Lattanzi}, {Lawrence}, {Leahy},
  {Leonardi}, {Lesgourgues}, {Levrier}, {Lewis}, {Liguori}, {Lilje},
  {Linden-V{\o}rnle}, {L{\'o}pez-Caniego}, {Lubin}, {Mac{\'\i}as-P{\'e}rez},
  {Maggio}, {Maino}, {Mandolesi}, {Mangilli}, {Marchini}, {Maris}, {Martin},
  {Martinelli}, {Mart{\'\i}nez-Gonz{\'a}lez}, {Masi}, {Matarrese}, {McGehee},
  {Meinhold}, {Melchiorri}, {Melin}, {Mendes}, {Mennella}, {Migliaccio},
  {Millea}, {Mitra}, {Miville-Desch{\^e}nes}, {Moneti}, {Montier}, {Morgante},
  {Mortlock}, {Moss}, {Munshi}, {Murphy}, {Naselsky}, {Nati}, {Natoli},
  {Netterfield}, {N{\o}rgaard-Nielsen}, {Noviello}, {Novikov}, {Novikov},
  {Oxborrow}, {Paci}, {Pagano}, {Pajot}, {Paladini}, {Paoletti}, {Partridge},
  {Pasian}, {Patanchon}, {Pearson}, {Perdereau}, {Perotto}, {Perrotta},
  {Pettorino}, {Piacentini}, {Piat}, {Pierpaoli}, {Pietrobon}, {Plaszczynski},
  {Pointecouteau}, {Polenta}, {Popa}, {Pratt}, {Pr{\'e}zeau}, {Prunet},
  {Puget}, {Rachen}, {Reach}, {Rebolo}, {Reinecke}, {Remazeilles}, {Renault},
  {Renzi}, {Ristorcelli}, {Rocha}, {Rosset}, {Rossetti}, {Roudier},
  {Rouill{\'e} d'Orfeuil}, {Rowan-Robinson}, {Rubi{\~n}o-Mart{\'\i}n},
  {Rusholme}, {Said}, {Salvatelli}, {Salvati}, {Sandri}, {Santos},
  {Savelainen}, {Savini}, {Scott}, {Seiffert}, {Serra}, {Shellard}, {Spencer},
  {Spinelli}, {Stolyarov}, {Stompor}, {Sudiwala}, {Sunyaev}, {Sutton},
  {Suur-Uski}, {Sygnet}, {Tauber}, {Terenzi}, {Toffolatti}, {Tomasi},
  {Tristram}, {Trombetti}, {Tucci}, {Tuovinen}, {T{\"u}rler}, {Umana},
  {Valenziano}, {Valiviita}, {Van Tent}, {Vielva}, {Villa}, {Wade}, {Wandelt},
  {Wehus}, {White}, {White}, {Wilkinson}, {Yvon}, {Zacchei}, \&
  {Zonca}}]{Planck15}
{Planck Collaboration}, {Ade}, P.~A.~R., {Aghanim}, N., {et~al.} 2016, \aap,
  594, A13, \dodoi{10.1051/0004-6361/201525830}

\bibitem[{Platts {et~al.}(2019)Platts, Weltman, Walters, Tendulkar, Gordin, \&
  Kandhai}]{Platts2019}
Platts, E., Weltman, A., Walters, A., {et~al.} 2019, Physics Reports, 821, 1,
  \dodoi{https://doi.org/10.1016/j.physrep.2019.06.003}

\bibitem[{Prochaska {et~al.}(2019)Prochaska, Simha, Law, Tejos, \&
  mneeleman}]{frbrepo}
Prochaska, J.~X., Simha, S., Law, C., Tejos, N., \& mneeleman. 2019, FRBs/FRB:
  First DOI release of this repository, v1.0.0,  Zenodo,
  \dodoi{10.5281/zenodo.3403651}

\bibitem[{Prochaska {et~al.}(2019b)Prochaska, Macquart, McQuinn, Simha,
  Shannon, Day, Marnoch, Ryder, Deller, Bannister, \& et~al.}]{Prochaska2019}
Prochaska, J.~X., Macquart, J.-P., McQuinn, M., {et~al.} 2019b, Science, 366,
  231–234, \dodoi{10.1126/science.aay0073}

\bibitem[{{Rajwade} {et~al.}(2022){Rajwade}, {Bezuidenhout}, {Caleb},
  {Driessen}, {Jankowski}, {Malenta}, {Morello}, {Sanidas}, {Stappers},
  {Surnis}, {Barr}, {Chen}, {Kramer}, {Wu}, {Buchner}, {Serylak}, {Combes},
  {Fong}, {Gupta}, {Jagannathan}, {Kilpatrick}, {Krogager}, {Noterdaeme},
  {N{\'u}nẽz}, {Prochaska}, {Srianand}, \& {Tejos}}]{Rajwade2022}
{Rajwade}, K.~M., {Bezuidenhout}, M.~C., {Caleb}, M., {et~al.} 2022, \mnras,
  514, 1961, \dodoi{10.1093/mnras/stac1450}

\bibitem[{Ravi {et~al.}(2019)}]{Ravi2019}
Ravi, V., {et~al.} 2019, Nature, 572, 352, \dodoi{10.1038/s41586-019-1389-7}

\bibitem[{{Roychowdhury} {et~al.}(2019){Roychowdhury}, {Arabsalmani}, \&
  {Kanekar}}]{Roychowdhury2019}
{Roychowdhury}, S., {Arabsalmani}, M., \& {Kanekar}, N. 2019, \mnras, 485, L93,
  \dodoi{10.1093/mnrasl/slz035}

\bibitem[{{Ryder} {et~al.}(2022){Ryder}, {Bannister}, {Bhandari}, {Deller},
  {Ekers}, {Glowacki}, {Gordon}, {Gourdji}, {James}, {Kilpatrick}, {Lu},
  {Marnoch}, {Moss}, {Prochaska}, {Qiu}, {Sadler}, {Simha}, {Sammons}, {Scott},
  {Tejos}, \& {Shannon}}]{Ryder2022}
{Ryder}, S.~D., {Bannister}, K.~W., {Bhandari}, S., {et~al.} 2022, arXiv
  e-prints, arXiv:2210.04680, \dodoi{10.48550/arXiv.2210.04680}

\bibitem[{{Saintonge} {et~al.}(2017){Saintonge}, {Catinella}, {Tacconi},
  {Kauffmann}, {Genzel}, {Cortese}, {Dav{\'e}}, {Fletcher},
  {Graci{\'a}-Carpio}, {Kramer}, {Heckman}, {Janowiecki}, {Lutz}, {Rosario},
  {Schiminovich}, {Schuster}, {Wang}, {Wuyts}, {Borthakur}, {Lamperti}, \&
  {Roberts-Borsani}}]{saintonge2017}
{Saintonge}, A., {Catinella}, B., {Tacconi}, L.~J., {et~al.} 2017, \apjs, 233,
  22, \dodoi{10.3847/1538-4365/aa97e0}

\bibitem[{{Shannon} {et~al.}(2018){Shannon}, {Macquart}, {Bannister}, {Ekers},
  {James}, {Os{\l}owski}, {Qiu}, {Sammons}, {Hotan}, {Voronkov}, {Beresford},
  {Brothers}, {Brown}, {Bunton}, {Chippendale}, {Haskins}, {Leach},
  {Marquarding}, {McConnell}, {Pilawa}, {Sadler}, {Troup}, {Tuthill},
  {Whiting}, {Allison}, {Anderson}, {Bell}, {Collier}, {G{\"u}rkan}, {Heald},
  \& {Riseley}}]{Shannonetal2018}
{Shannon}, R.~M., {Macquart}, J.~P., {Bannister}, K.~W., {et~al.} 2018, \nat,
  562, 386, \dodoi{10.1038/s41586-018-0588-y}

\bibitem[{{Sharma} {et~al.}(2023){Sharma}, {Somalwar}, {Law}, {Ravi}, {Catha},
  {Chen}, {Connor}, {Faber}, {Hallinan}, {Harnach}, {Hellbourg}, {Hobbs},
  {Hodge}, {Hodges}, {Lamb}, {Rasmussen}, {Sherman}, {Shi}, {Simard},
  {Squillace}, {Weinreb}, {Woody}, \& {Yadlapalli}}]{Sharma2023}
{Sharma}, K., {Somalwar}, J., {Law}, C., {et~al.} 2023, arXiv e-prints,
  arXiv:2302.14782, \dodoi{10.48550/arXiv.2302.14782}

\bibitem[{{Simha} {et~al.}(2021){Simha}, {Tejos}, {Prochaska}, {Lee}, {Ryder},
  {Cantalupo}, {Bannister}, {Bhandari}, \& {Shannon}}]{Simha2021}
{Simha}, S., {Tejos}, N., {Prochaska}, J.~X., {et~al.} 2021, \apj, 921, 134,
  \dodoi{10.3847/1538-4357/ac2000}

\bibitem[{{Solomon} \& {Vanden Bout}(2005)}]{solomon2005}
{Solomon}, P.~M., \& {Vanden Bout}, P.~A. 2005, \araa, 43, 677,
  \dodoi{10.1146/annurev.astro.43.051804.102221}

\bibitem[{{Speagle} {et~al.}(2014){Speagle}, {Steinhardt}, {Capak}, \&
  {Silverman}}]{Speagle2014}
{Speagle}, J.~S., {Steinhardt}, C.~L., {Capak}, P.~L., \& {Silverman}, J.~D.
  2014, \apjs, 214, 15, \dodoi{10.1088/0067-0049/214/2/15}

\bibitem[{{Spitler} {et~al.}(2016){Spitler}, {Scholz}, {Hessels}, {Bogdanov},
  {Brazier}, {Camilo}, {Chatterjee}, {Cordes}, {Crawford}, {Deneva}, {Ferdman},
  {Freire}, {Kaspi}, {Lazarus}, {Lynch}, {Madsen}, {McLaughlin}, {Patel},
  {Ransom}, {Seymour}, {Stairs}, {Stappers}, {van Leeuwen}, \&
  {Zhu}}]{Spitler2016}
{Spitler}, L.~G., {Scholz}, P., {Hessels}, J.~W.~T., {et~al.} 2016, \nat, 531,
  202, \dodoi{10.1038/nature17168}

\bibitem[{{Tacconi} {et~al.}(2018){Tacconi}, {Genzel}, {Saintonge}, {Combes},
  {Garc{\'\i}a-Burillo}, {Neri}, {Bolatto}, {Contini}, {F{\"o}rster Schreiber},
  {Lilly}, {Lutz}, {Wuyts}, {Accurso}, {Boissier}, {Boone}, {Bouch{\'e}},
  {Bournaud}, {Burkert}, {Carollo}, {Cooper}, {Cox}, {Feruglio}, {Freundlich},
  {Herrera-Camus}, {Juneau}, {Lippa}, {Naab}, {Renzini}, {Salome}, {Sternberg},
  {Tadaki}, {{\"U}bler}, {Walter}, {Weiner}, \& {Weiss}}]{Tacconi2018}
{Tacconi}, L.~J., {Genzel}, R., {Saintonge}, A., {et~al.} 2018, \apj, 853, 179,
  \dodoi{10.3847/1538-4357/aaa4b4}

\bibitem[{Tendulkar {et~al.}(2017)Tendulkar, Bassa, Cordes, Bower, Law,
  Chatterjee, Adams, Bogdanov, Burke-Spolaor, Butler, Demorest, Hessels, Kaspi,
  Lazio, Maddox, Marcote, McLaughlin, Paragi, Ransom, Scholz, Seymour, Spitler,
  van Langevelde, \& Wharton}]{Tendulkar2017}
Tendulkar, S.~P., Bassa, C.~G., Cordes, J.~M., {et~al.} 2017, The Astrophysical
  Journal Letters, 834, L7, \dodoi{10.3847/2041-8213/834/2/L7}

\bibitem[{{Tendulkar} {et~al.}(2021){Tendulkar}, {Gil de Paz}, {Kirichenko},
  {Hessels}, {Bhardwaj}, {{\'A}vila}, {Bassa}, {Chawla}, {Fonseca}, {Kaspi},
  {Keimpema}, {Kirsten}, {Lazio}, {Marcote}, {Masui}, {Nimmo}, {Paragi},
  {Rahman}, {Pay{\'a}}, {Scholz}, \& {Stairs}}]{Tendulkar2021}
{Tendulkar}, S.~P., {Gil de Paz}, A., {Kirichenko}, A.~Y., {et~al.} 2021,
  \apjl, 908, L12, \dodoi{10.3847/2041-8213/abdb38}

\bibitem[{{Virtanen} {et~al.}(2020){Virtanen}, {Gommers}, {Oliphant},
  {Haberland}, {Reddy}, {Cournapeau}, {Burovski}, {Peterson}, {Weckesser},
  {Bright}, {van der Walt}, {Brett}, {Wilson}, {Millman}, {Mayorov}, {Nelson},
  {Jones}, {Kern}, {Larson}, {Carey}, {Polat}, {Feng}, {Moore}, {VanderPlas},
  {Laxalde}, {Perktold}, {Cimrman}, {Henriksen}, {Quintero}, {Harris},
  {Archibald}, {Ribeiro}, {Pedregosa}, {van Mulbregt}, \& {SciPy 1. 0
  Contributors}}]{virtanen2020}
{Virtanen}, P., {Gommers}, R., {Oliphant}, T.~E., {et~al.} 2020, Nature
  Methods, 17, 261, \dodoi{10.1038/s41592-019-0686-2}

\end{thebibliography}

\end{document}